\documentclass[12pt]{iopart}
\pdfoutput=1
\usepackage[applemac]{inputenc}
\usepackage[english]{babel}
\usepackage[pdftex]{graphicx}
\usepackage{color}
\usepackage{multirow}

\begin{document}

\title[Thermal and Quantum Phase Transitions in Atom-Field Systems]{Thermal and Quantum Phase Transitions in Atom-Field Systems: \\ a Microcanonical Analysis.}
\author{M. A. Bastarrachea-Magnani} 
\address{Instituto de Ciencias Nucleares, Universidad Nacional Aut\'onoma de M\'exico,
Apdo. Postal 70-543, M\'exico D. F., C.P. 04510}
\author{S. Lerma-Hern\'andez}
\address{Facultad  de F\'\i sica, Universidad Veracruzana,
Circuito Aguirre Beltr\'an s/n, Xalapa, Veracruz, M\'exico, C.P. 91000}
\author{J. G. Hirsch}
\address{Instituto de Ciencias Nucleares, Universidad Nacional Aut\'onoma de M\'exico,
Apdo. Postal 70-543, M\'exico D. F., C.P. 04510}


\begin{abstract}
The thermodynamical properties of a generalized Dicke model are calculated and related with the critical properties of its energy spectrum, namely the quantum phase transitions (QPT) and excited state quantum phase transitions (ESQPT). The thermal properties are calculated both in the canonical and the microcanonical ensembles. The latter deduction allows for an explicit description of the relation between thermal and energy spectrum properties. While in an isolated system the subspaces with different pseudo spin are disconnected, and the whole energy spectrum is accesible, in the thermal ensamble the situation is radically different. The multiplicity  of the lowest energy states for each pseudo spin completely dominates the thermal behavior, making the set of degenerate states with the smallest pseudo spin at a given energy the only ones playing a role in the thermal properties, making the positive energy states thermally inaccesible. Their quantum phase transitions, from a normal to a superradiant  phase, are closely associated with the thermal transition. The other critical phenomena, the ESQPTs occurring at excited energies, have no manifestation in the thermodynamics, although their effects could be seen in finite sizes corrections. A new superradiant phase is found, which only exists in the generalized model,  and can be relevant in finite size systems.   
\end{abstract}

\vspace{2pc}

\maketitle


\section{Introduction}

The description of non-equilibrium dynamics and thermalization in isolated quantum many-body systems has received renewed interest during the last years thanks mainly  to the development of novel and powerful numerical approaches to study these problems, as well as, to the advance in sophisticated experimental techniques to control quantum systems with many degrees of freedom \cite{Pol11,Eis15,Gog15}. While the connection between thermodynamics and statistical physics is clear in the context of microscopic laws governed by classical mechanics, for quantum mechanics is not the case because of the difficulties in defining important concepts like chaos, integrability and analytic solvability. We lack a complete framework able to explain how equilibrium and thermal states described by the ensembles of statistical mechanics arise from a microscopic description, and how a quantum phase transition (QPT) and its dynamics could be interpreted from the thermodynamic point of view. 

Furthermore, in the case of many particles and degrees of freedom, the techniques developed to deal with the zero temperature case cannot easily be extended for solving finite temperature problems \cite{Wilms12}. At zero temperature, quantum systems occupy only the ground-state. However, with finite temperature, a quantum system has enough thermal energy to occupy excited states. Therefore, in order to study finite temperature problems in quantum many-body systems the information of their spectra is significant. This leads us to relate the point of view of statistical ensambles, where the temperature plays an important role, with the perspective of isolated systems, where the spectrum is relevant. 

One interesting feature of the spectrum in quantum many-body systems is the excited-state quantum phase transition (ESQPT). An ESQPT is a singularity in the density of states which takes place along the energy spectrum for fixed values of the Hamiltonian parameters \cite{Cap08} and has a strong semi-classical connection \cite{Str15,Cej15}. The ESQPTs have been studied in several nuclear physics models \cite{Cej06} and it has been suggested they could have important effects in decoherence \cite{Rel09} and the temporal evolution of quantum quenches \cite{Per11,San15}. The relationship between the ground state QPTs and ESQPTs is not completely clear, neither are the dynamical properties of the latter, so these issues are open to current research.

The aim of this work is to relate some critical features of the quantum spectrum in atom-field systems, specifically the QPT and ESQPT, with their thermodynamics. We study these features in a generalized Dicke model, including  the Dicke and Tavis-Cummings Hamiltonians \cite{Dicke54,TC68}, which describe a system of $\mathcal{N}$ two-level atoms interacting with a single monochromatic electromagnetic radiation mode within a cavity. In the language of quantum computing and quantum information, they also describe a set of $\mathcal{N}$ qubits from quantum dots, Bose-Einstein condensates in optical lattices, and circuit QED \cite{Bau10,Nag10,Sch03,Sch07,Bla04,Fink09} interacting through a bosonic mode. 

Both Hamiltonians are paradigmatic examples of quantum collective behavior in quantum optics. While the Dicke Hamiltonian is non-integrable, the Tavis-Cummings Hamiltonian is its integrable version due the rotating-wave approximation (RWA). The Dicke model is interesting not only because its experimental realizations, but also thanks to its critical phenomena: the superradiant thermal phase transition \cite{HL73}, the well-known superradiant QPT, related with quantum chaos and entanglement \cite{Bran03}, and the presence of (\emph{dynamic} and \emph{static}) ESQPTs \cite{Per112,Bran13,Bas14,Bas15}. The Dicke model is a suited toy model to explore the connection between thermodynamics and the spectrum of quantum systems. In order to address both Hamiltonians at once, and at the same time to have the possibility to go from integrability to non-integrability, we put together the Dicke and Tavis-Cummings Hamiltonians in one expression introducing a control parameter. 
We call it the generalized Dicke Hamiltonian. 

Originally, the thermodynamic analysis for the Tavis-Cummings model was presented by K. Hepp and E. H. Lieb in 1973 \cite{HL73}. Their method was simplified trough the Laplace's integral method and extended for the multimode case by Wang and Hioe \cite{WH73,Hioe73}. Later, the counter-rotating terms were included in \cite{CGW73,CD74}. In the following years, several authors developed different methods and approaches to study the thermodynamics of the Dicke model \cite{Gibb74,Bra75,Gil76, Lee76,Ver74,Ors771,Ors772, Apa12, Gil81, Lib05}. All these approaches rely on the canonical ensemble. The only analysis in the microcanonical ensamble we are aware of, presents some partial results employing non-normalized Gaussian distributions  \cite{Jaw85}.

In this work, we calculate the thermodynamic properties of the generalized Dicke model. After a review of the well-known procedure to calculate the canonical partition function employing the Laplace's integral method, used as reference, we make the calculation in the micro-canonical ensemble, building a natural link between the thermodynamics and the properties of the quantum spectrum. 

As the generalized Dicke Hamiltonian commutes with the total pseudo spin operator $\vec{\mathbf{J}}^{2}$, the Hamiltonian is block-diagonal in subspaces labeled with $j$ running from $0$ to $\mathcal{N}/2$, where $j(j+1)$ is the eigenvalue of the total pseudo spin operator. In the last years most of the studies about the Dicke model and its QPT have been restricted to the symmetric representation i. e. the subspace with maximum pseudo-spin sector $j_M=\mathcal{N}/2$, where the ground-state lies. Nevertheless, in order to describe the thermodynamic properties of the full spectrum it is necessary to include all the $j$ sectors. As mentioned above, a satisfactory framework for this is still missing, so we employ a semi-classical approach to calculate the microcanonical ensamble. 

This article is organized as follows. In section 2 we calculate the thermodynamics of the generalized Dicke model in the canonical ensemble. The results of this section are recovered in section 3, but from a microcanonical approach. In order to obtain the number of states for a given energy,  a semi-classical approximation to the ground-state  energies and to the density of states is obtained for  each sector $0\leq j \leq \mathcal{N}/2$. Likewise, the thermodynamical limit of the multiplicities $Y(\mathcal{N},j)$ is discussed and obtained.  Finally, we give our conclusions. Besides, we present several Appendixes with a detailed discussion of the calculations and other considerations about the thermal phases.   


\section{Canonical thermodynamics of generalized Dicke model}
In this section, we calculate the thermodynamics of the generalized Dicke model following the traditional procedure for calculating the canonical partition function \cite{WH73,Hioe73,CGW73,CD74}. The generalized Dicke Hamiltonian is,
\begin{equation}
H_{D,\delta}=\omega a^{\dagger}a+ \omega_{0} J_{z} + \frac{\gamma}{\sqrt{\mathcal{N}}}\left[\frac{}{}(1+\delta)(a+a^{\dagger}) J_{x} - i(1-\delta) (a^{\dagger}-a) J_{y} \right].
\end{equation}
With $\delta\in[0,1]$. When $\delta=0$ we recover the integrable Tavis-Cummings Hamiltonian \cite{TC68}, meanwhile when $\delta=1$ we have the non-integrable Dicke Hamiltonian \cite{Dicke54}. The pseudo-spin collective operators are defined in terms of the Pauli operators as $J_{\mu}=\frac{1}{2}\sum_{k=0}^{\mathcal{N}}\sigma_{\mu}^{k}$ (with $\mu=x,y,z$). With this we can write the Hamiltonian as, 
\begin{eqnarray}
\fl H_{D,\delta}&=\sum_{k=1}^{\mathcal{N}} H_{D,\delta}^{k}=\\  
\fl &=\sum_{k=1}^{\mathcal{N}}\left\{ \omega \frac{a}{\sqrt{\mathcal{N}}} \frac{a^{\dagger}}{\sqrt{\mathcal{N}}} + \frac{\omega_{0}}{2} \sigma_{z}^{k}+\frac{\gamma}{2\sqrt{\mathcal{N}}} \left[(1+\delta)(a+a^{\dagger}) \sigma_{x}^{k} - i(1-\delta) (a^{\dagger}-a) \sigma_{y}^{k} \right)\right\}. \nonumber
\end{eqnarray} 
We want to calculate the canonical partition function,
\begin{equation}
\mathcal{Z}_{\delta}(T,\mathcal{N})={\hbox {Tr}}\left(e^{-\beta\,H_{D,\delta}}\right),
\end{equation}
with $\beta = 1/k_B T$, being $k_B$ the Boltzmann's constant.
In order to obtain the trace, we chose Glauber coherent states for the field part and single pseudo-spin states for the atomic sector, 
\begin{equation}
|\Psi\rangle=|\alpha\rangle\otimes |s_{1}\rangle\otimes|s_{2}\rangle\otimes\cdot\cdot\cdot\otimes|s_{\mathcal{N}}\rangle.
\end{equation}
We calculate the canonical partition function as 
\begin{equation}
 \label{zfunc}
\fl \mathcal{Z}_{\delta}(T,\mathcal{N})=\int \frac{d^{2}\alpha}{\pi}\sum_{s_{1}=\pm}\sum_{s_{2}=\pm}\cdot\cdot\cdot\sum_{s_{\mathcal{N}}=\pm} \langle\alpha|\langle s_{1}|\langle s_{2}|\cdot\cdot\cdot\langle s_{\mathcal{N}}| e^{-\beta H_{D,\delta}} |\alpha\rangle |s_{1}\rangle|s_{2}\rangle\cdot\cdot\cdot|s_{\mathcal{N}}\rangle.
\end{equation}
The result is (for details see Appendix A)
\begin{equation}
\mathcal{Z}_{\delta}(T,\mathcal{N})=\frac{\mathcal{N}}{\pi}\int_{-\infty}^{\infty}\int_{-\infty}^{\infty}du_{+}\,du_{-}\, \,e^{\,\mathcal{N} \phi_{\delta}(u_{+},u_{-})},
\end{equation}
where $\frac{\alpha}{\sqrt{\mathcal{N}}}=u_{+}+iu_{-}$, and, for convenience, we have defined  new functions
\begin{equation}
\phi_{\delta}(u_{+},u_{-})= -\beta\omega(u_{+}^{2}+u_{-}^{2}) + \ln\left\{2\, \cosh\, \left[\frac{\beta\omega_{0}}{2}\, \chi_{\delta}(u_{+},u_{-})\right]\right\}
\end{equation}
with 
\begin{equation}
\chi_{\delta}(u_{+},u_{-})=\sqrt{1+\frac{4\gamma^{2}}{\omega_{0}^{2}}\left[(1+\delta)^{2} u_{+}^{2}+(1-\delta)^{2} u_{-}^{2}\right]}.
\label{chi_delta}
\end{equation}
The free energy $\mathcal{F}$, entropy $\mathcal{S}$, and energy $\mathcal{U}$ per particle are
\begin{eqnarray}
-\beta\mathcal{F}_{\delta}(T)&=\lim_{\mathcal{N}\rightarrow\infty} \frac{1}{\mathcal{N}}\,\ln\left[\frac{}{}\mathcal{Z}_{\delta}\left(T,\mathcal{N}\right)\right],\\
\frac{\mathcal{S}_{\delta}(T)}{k_{B}}&=\lim_{\mathcal{N}\rightarrow\infty} \frac{1}{\mathcal{N}} \left\{\ln\left[\mathcal{Z}_{\delta}(T,\mathcal{N})\right]+\frac{1}{k_{B}\beta\mathcal{Z}}\frac{\partial \mathcal{Z}_{\delta}(T,\mathcal{N})}{\partial T}\right\},\\
\mathcal{U}_{\delta}(T)&=\lim_{\mathcal{N}\rightarrow\infty} \frac{1}{\mathcal{N}}\left[\frac{}{}F(T,\mathcal{N})+T\,S(T,\mathcal{N})\right].
\end{eqnarray}


\subsection{Thermal Averages of $a^{\dagger}a$ and $J_{\mu}$}

To calculate the thermodynamic expectation values of some observables of interest like the number of photons and the pseudo-spin collective operators, we proceed as follows. 

\subsubsection{Number operator.}

To calculate the number operator and its powers,  we use the same formalism employed to find the partition function. Because the number operator only affects the boson trace we have,  
\begin{equation}
\left\langle\left(\frac{a^{\dagger}a}{\mathcal{N}}\right)^{k}\right\rangle_{\delta}=\frac{1}{\mathcal{Z}_{\delta}}\int \frac{d^{2}\alpha}{\pi} \left(\frac{|\alpha|^{2}}{\mathcal{N}}\right)^{k}\,e^{\mathcal{N}\,\phi_{\delta}(\alpha)}.
\end{equation}
Employing the variables $u_{\pm}$ and expanding the Newton binomial we have the final expression,  
\begin{equation} \label{ch3eqnumfot}
\fl \left\langle\left(\frac{a^{\dagger}a}{\mathcal{N}}\right)^{k}\right\rangle_{\delta}=\frac{\mathcal{N}}{\mathcal{Z}_{\delta}} \sum_{l=0}^{k}\left(\begin{array}{c} k \\ l \\ \end{array}\right)\int_{-\infty}^{\infty}\int_{-\infty}^{\infty} \frac{du_{+}\,du_{-}}{\pi} (u_{+}^{2})^{k-l}(u_{-}^{2})^{l}\,e^{\mathcal{N}\,\phi_{\delta}(u_{+},u_{-})}.
\end{equation}


\subsubsection{Collective pseudo-spin operators.}

Now, we find an expression for the thermal average of the collective pseudo-spin operators $J_{\mu}$ with $\mu=x,y,z$. In this case, the operator does not affect the photon trace, only the atomic one, so we come back a few steps in order to calculate it. We start from the photon trace, 
\begin{eqnarray}
\fl & \left\langle\frac{J_{\mu}}{\mathcal{N}}\right\rangle_{\delta}=\\ \nonumber
\fl &=\frac{1}{\mathcal{N}\mathcal{Z}_{\delta}}\int \frac{d^{2}\alpha}{\pi}\,e^{-\beta\omega|\alpha|^{2}}\sum_{s_{1}=\pm}\sum_{s_{2}=\pm}\cdot\cdot\cdot\sum_{s_{\mathcal{N}}=\pm}\langle s_{1}|\langle s_{2}|\cdot\cdot\cdot\langle s_{\mathcal{N}}| J_{\mu}e^{-\beta\,\sum_{k}h_{k}(\alpha)} |s_{\mathcal{N}}\rangle\cdot\cdot\cdot|s_{2}\rangle|s_{1}\rangle.
\end{eqnarray}

Using that $J_{\mu}=\frac{1}{2}\sum_{\ell}^{\mathcal{N}}\sigma_{\mu}^{\ell}$, we see it is possible to reorder the index to obtain the following expression
\begin{equation}
\left\langle\frac{J_{\mu}}{\mathcal{N}}\right\rangle=\frac{1}{\mathcal{N}\mathcal{Z}_{\delta}}\sum_{\ell=1}^{\mathcal{N}}\int \frac{d^{2}\alpha}{2\pi}\,e^{-\beta\omega|\alpha|^{2}}\ \prod_{k}^{\mathcal{N}}\sum_{s_{k}=\pm}\langle s_{k}| \sigma_{\mu}^{\ell}\,e^{-\beta h_{k}(\alpha)}|s_{k}\rangle.
\end{equation}
The final result is (for details see Appendix B) 
\begin{eqnarray} \label{eq:jzxy}
\nonumber
\left\langle\frac{J_{\mu}}{\mathcal{N}}\right\rangle_{\delta}&=-\frac{\mathcal{N}}{\mathcal{Z}_{\delta}}\int_{-\infty}^{\infty}\int_{-\infty}^{\infty} \frac{du_{+}\,du_{-}}{2\pi}\,\frac{\,\tanh\left(\frac{\beta\omega_{0}}{2}\chi_{\delta}(u_{+},u_{-})\right)}{\omega_{0}\,\chi_{\delta}(u_{+},u_{-})}\,e^{\mathcal{N}\phi_{\delta}(u_{+},u_{-})}\times\\ 
&\times\left\{\omega_{0}\delta_{\mu,z}+\gamma\left[(1+\delta)\,u_{+}\,\delta_{\mu,x}-(1-\delta)\,u_{-}\,\delta_{\mu,y}\right]\right\}.
\end{eqnarray}


\subsection{Solving the canonical partition function}

The form of the partition function is specially suitable for using the steepest descents method or Laplace's integral method \cite{Orz78} to calculate it. This method consists in approximating 
the exponential integrand by a gaussian function around the global maximum of the function $\phi_\delta(u_+,u_-)$
$$
\mathcal{Z}_\delta(T,\mathcal{N})=\frac{\mathcal{N}}{\pi}\int_\infty^\infty du_+ du_- e^ {\mathcal{N}\phi_\delta(u_+,u_)}\approx \frac{2}{\sqrt{D(u_+^m,u_-^m)}}\exp(\mathcal{N}\phi_\delta(u_+^m,u_-^m)),
$$
where $u_\pm^m$ are the extremal values of
$\phi_\delta(u_+,u_-)$, and $D(u_+,u_-)=\partial^{2}_{u_{-}}\phi_{\delta}\partial^{2}_{u_{+}}\phi_{\delta}-\left(\partial_{u_{+}u_{-}}\phi_{\delta}\right)^2$ is the determinant of the Hessian matrix. 

In order to calculate the extremal values of $\phi_{\delta}(u_{+},u_{-})$, 
 we  obtain the derivatives of $\chi_{\delta}(u_{+},u_{-})$, 
\begin{eqnarray}
\frac{\partial \chi_{\delta}}{\partial u_{\pm}}&=\frac{4\gamma^{2}}{\omega_{0}^{2}}\left(1\pm\delta\right)^{2}\frac{u_{\pm}}{\chi_{\delta}(u_{+},u_{-})}=\frac{4\omega}{\omega_{0}}\left(\frac{\gamma}{\gamma_{\pm}}\right)^{2}\frac{u_{\pm}}{\chi_{\delta}(u_{+},u_{-})},\\
\end{eqnarray}
where we have defined 
\begin{equation}
\gamma_{\pm}=\frac{\sqrt{\omega\omega_{0}}}{1\pm\delta}
\end{equation}

Therefore, the first derivatives of $\phi_{\delta}(u_{+},u_{-})$ are
\begin{eqnarray} \label{eq1}
\frac{\partial \phi_{\delta}}{\partial u_{\pm}}&= -2u_{\pm}\beta\omega\left\{1 - \left(\frac{\gamma}{\gamma_{\pm}}\right)^{2}\frac{1}{\chi_{\delta}(u_{+},u_{-})}\tanh\left[\frac{}{}\eta\, \chi_{\delta}(u_{+},u_{-})\right]\right\},
\end{eqnarray}
with \begin{equation}\eta=\beta\omega_{0}/2.\end{equation} 

The second derivatives are 
\begin{eqnarray}
\fl \frac{\partial^{2} \phi_{\delta}}{\partial u_{\pm}^{2}}&= -2\beta\omega\left\{1 - \left(\frac{\gamma}{\gamma_{\pm}}\right)^{2}\frac{1}{\chi_{\delta}(u_{+},u_{-})}\tanh\left[\frac{}{}\eta\, \chi_{\delta}(u_{+},u_{-})\right]\right.\\ \nonumber
\fl &- \left.\frac{4\omega}{\omega_{0}}\left(\frac{\gamma}{\gamma_{\pm}}\right)^{4}\frac{u_{\pm}^{2}}{\chi^{3}_{\delta}(u_{+},u_{-})}\left[\eta\,\chi_{\delta}(u_{+},u_{-})\,\mbox{sech}^{2}\left[\frac{}{}\eta\, \chi_{\delta}(u_{+},u_{-})\right]+\right.\right. \\ \nonumber
\fl &\left.\left. -\tanh\left[\frac{}{}\eta\, \chi_{\delta}(u_{+},u_{-})\right]\right]\right\},\\
\fl \frac{\partial^{2} \phi_{\delta}}{\partial u_{+}\,\partial u_{-}}&=-2\beta\omega \frac{u_{+}\,u_{-}}{\chi^{3}_{\delta}(u_{+},u_{-})}\frac{4\omega}{\omega_{0}}\left(\frac{\gamma}{\gamma_{+}}\right)^{2}\left(\frac{\gamma}{\gamma_{-}}\right)^{2}\times \\ \nonumber
\fl &\times \left\{\tanh\left[\frac{}{}\eta\, \chi_{\delta}(u_{+},u_{-})\right]-\eta\,\chi_{\delta}(u_{+},u_{-})\,\mbox{sech}^{2}\left[\frac{}{}\eta\, \chi_{\delta}(u_{+},u_{-})\right]\right\}.
\end{eqnarray}

If we look for the extremal points  by making $\nabla \phi_{\delta}(u^{m}_{+},u^{m}_{-})=0$, we have three possibilities {\bf {a)}} both $u^{m}_{\pm}=0$, {\bf{b)}} $u^{m}_{+}\neq 0$ and $u_{-}^{m}= 0$, {\bf c)}  $u_{-}^{m}\neq 0$ and $u_{+}^{m}= 0$.  The  case   $u_{\pm}^{m}\neq0$ implies that the two equations  ($\pm$)
\begin{equation}
\tanh\left(\eta\,\chi_{\delta}\right)=\left(\frac{\gamma_{\pm}}{\gamma}\right)^{2}\chi_{\delta},
\end{equation} 
must hold simultaneously for the same function $\chi_{\delta}(u_{+},u_{-})$, which is impossible, therefore there is not  a maximum with both $u_{\pm}^{m}\neq0$ and this case can be discarded. 

In the following we analyze the remaining  cases separately. 


\subsubsection{Normal phase $u_{\pm}^{m}=0$.}

This  first case,  $u_{\pm}^{m}=0$, corresponds  clearly to  a \emph{normal phase} due to the fact $\mathcal{N}|\alpha|^{2}=u_{+}^{2}+u_{-}^{2}=0$, i.e. there is an average of zero photons in the field. By noting that $\chi_{\delta}(0,0)=1$, the function and its second derivatives become,
\begin{eqnarray}
&\phi_{\delta}(0,0)= \ln\left\{2\, \cosh\, \left(\eta\right)\right\},\\ \nonumber
&\left.\frac{\partial^{2} \phi_{\delta}}{\partial u_{\pm}^{2}}\right|_{(0,0)}= -2\beta\omega\left\{1 - \left(\frac{\gamma}{\gamma_{\pm}}\right)^{2}\tanh\left[\eta\right]\right\},\\ \nonumber
&\left.\frac{\partial^{2} \phi_{\delta}}{\partial u_{+}\,\partial u_{-}}\right|_{(0,0)}=0.
\end{eqnarray}
Therefore, the determinant of the hessian matrix is 
\begin{eqnarray}
D(0,0)=4\beta^{2}\omega^{2}\left\{1 - \left(\frac{\gamma}{\gamma_{+}}\right)^{2}\,\tanh\left[\eta\right]\right\}\left\{1 - \left(\frac{\gamma}{\gamma_{-}}\right)^{2}\,\tanh\left[\eta\right]\right\}.
\end{eqnarray}
To ensure that this point corresponds to a maximum, and consequently that the Laplace's method around this point can be used to approximate the partition function, the conditions $D(0,0)>0$ and $\left.\frac{\partial^{2} \phi_{\delta}}{\partial u_{+}^{2}}\right|_{(0,0)}<0$  must hold simultaneously. Since $\gamma_-\geq \gamma_+$, these conditions are satisfied if and only if 
\begin{equation}
\gamma^2\tanh\eta<\gamma_+^2.
\label{condNormal}
\end{equation}
The previous condition defines, consequently, the \emph{normal phase}  region in the parameters-Temperature ($\delta-\gamma-T$) space. Observe that, for $\gamma<\gamma_+$ the previous condition is satisfied for any value of the temperature, whereas for  $\gamma>\gamma_+$ the condition defines a range $\beta\in [0,\beta_c^{+}]$ ($T\in [T_c^{+},\infty]$ in terms of Temperature) for  the normal phase, where 
\begin{equation}
\beta_c^{+}=\frac{2}{\omega_o}\mbox{arctanh}\left(\frac{\gamma_+}{\gamma}\right)^2, \ \ \ \gamma>\gamma_+
\label{betacrit}
\end{equation}
  
Using the  Laplace's method around the point $(u_+,u_-)=(0,0)$  gives 
\begin{equation}
\mathcal{Z}_{\delta}(T,\mathcal{N})=\mathcal{O}_{\delta}(\mathcal{N})\,\exp\left\{\mathcal{N}\ln\left[2 \cosh(\eta)\right]\right\},
\end{equation}
where $\mathcal{O}_{\delta}(\mathcal{N})$ is a function of order $\mathcal{N}^0$ which is  negligible   in the thermodynamic limit
\begin{equation}
\mathcal{O}_{\delta}(\mathcal{N})=\frac{1}{\beta\omega}\left(1 - \left(\frac{\gamma}{\gamma_{+}}\right)^{2}\tanh\left[\eta\right]\right)^{-1/2}\left(1 - \left(\frac{\gamma}{\gamma_{-}}\right)^{2}\tanh\left[\eta\right]\right)^{-1/2}.
\end{equation}

The free energy per particle is
\begin{equation}
-\beta \mathcal{F}_\delta(T)=\ln\left[2 \cosh(\eta)\right].
\end{equation}
We calculate the partition function's first derivative, 
\begin{equation}
\frac{1}{\beta\mathcal{Z}_{\delta}}\frac{\partial \mathcal{Z}_{\delta}}{\partial T}=\frac{1}{\beta\mathcal{O}}\frac{\partial \mathcal{O}}{\partial T}-k_{B}\,\mathcal{N}\eta\,\tanh\left(\eta\right), 
\end{equation}
to obtain  the entropy per particle
\begin{equation}
\frac{\mathcal{S}_{\delta}(T)}{k_{B}}=\frac{\ln \mathcal{Z}_\delta}{\mathcal{N}}+\frac{1}{k_B \mathcal{N}}\frac{1}{\beta\mathcal{Z}_{\delta}}\frac{\partial \mathcal{Z}_{\delta}}{\partial T}= \ln\left[2\frac{}{}\cosh(\eta)\right]-\eta\,\tanh(\eta).
\end{equation} 
Therefore  the energy per particle is 
\begin{equation}
\mathcal{U}_{\delta}(T)= \mathcal{F}(T)+T \mathcal{S}_{\delta}(T)= -\frac{\omega_{0}}{2}\,\tanh(\eta).
\end{equation} 
Also, we can calculate the heat capacity, which is
\begin{equation}
C_{\delta}=\frac{\partial \mathcal{U}_{\delta}}{\partial T}=k_{B}\eta^{2}\,\mbox{sech}^{2}\left(\eta\right).
\end{equation}

Similarly, by using the Laplace's method, after both integrations, the thermal average of the photon number is
\begin{equation}
n_{\delta}^{k}=\lim_{\mathcal{N}\rightarrow\infty}\left\langle\left(\frac{a^{\dagger}a}{\mathcal{N}}\right)^{k}\right\rangle_{\delta}=0,
\end{equation}
and the thermal average of the collective pseudo-spin operator is
\begin{eqnarray}
\fl \sigma_{\delta,\mu}&=\lim_{\mathcal{N}\rightarrow\infty}\left\langle\frac{J_{\mu}}{\mathcal{N}}\right\rangle_{\delta}=\\ \nonumber
\fl& =\left.-\frac{\,tanh\left(\frac{\beta\omega_{0}}{2}\chi_{\delta}(u_{+},u_{-})\right)}{2\omega_{0}\,\chi_{\delta}(u_{+},u_{-})}\,\left\{\omega_{0}\delta_{\mu,z}+\gamma\left[(1+\delta)\,u_{+}\,\delta_{\mu,x}-(1-\delta)\,u_{-}\,\delta_{\mu,y}\right]\right\}\right|_{0,0}=\\ \nonumber
\fl \sigma_{\delta,\mu}&=-\frac{1}{2}\tanh(\eta)\delta_{\mu,z}.
\end{eqnarray}

Finally, as we are interested on the energy diagram, we can express entropy in terms of energy. We note that
\begin{equation}
\eta=\,\mbox{arctanh}\left(-\frac{2\mathcal{U}_{\delta}}{\omega_{0}}\right).
\end{equation}
Then
\begin{equation}
\frac{\mathcal{S}(\mathcal{U}_{\delta})}{k_{B}}=\ln(2)-\frac{1}{2}\ln\left[1-\left(-\frac{2\mathcal{U}_{\delta}}{\omega_{0}}\right)^{2}\right]+\left(-\frac{2\mathcal{U}_{\delta}}{\omega_{0}}\right)\,\mbox{arctanh}\left(-\frac{2\mathcal{U}_{\delta}}{\omega_{0}}\right).
\end{equation}
If we define
\begin{equation}
\mathcal{E}_{\delta}^{n.ph.}=-\frac{2\mathcal{U}_{\delta}}{\omega_{0}},
\end{equation}
we finally have 
\begin{equation}
\frac{\mathcal{S}(\mathcal{E}_{\delta}^{n.ph.})}{k_{B}}=\ln(2)-\frac{1}{2}\ln\left[1-\left(\mathcal{E}_{\delta}^{n.ph.}\right)^{2}\right]-\mathcal{E}_{\delta}^{n.ph.}\,\mbox{arctanh}\left(\mathcal{E}_{\delta}^{n.ph.}\right).
\end{equation}

Next, we calculate the remaining cases. 


\subsubsection{The two Dicke superradiant phases.} 

The second and third cases corresponds to consider one $u_{\pm}=u_{\pm}^{m}\neq0$ and the other $u_{\mp}^{m}=0$. These cases are related to a \emph{superradiant phase} due to the fact $\mathcal{N}|\alpha|^{2}=u_{\pm}^{2}\neq0$.  The condition for having an extreme point  (\ref{eq1}) for 
$u_{\pm}\neq0$ and $u_\mp=0$, reduces to
\begin{equation}
\label{eq:sola}
\tanh\left(\eta\,\chi_{\delta}\right)=\left(\frac{\gamma_{\pm}}{\gamma}\right)^{2}\chi_{\delta}.
\end{equation}
Because of the definition of $\chi_{\delta}$, Eq.(\ref{chi_delta}), we are only  interested in solutions of the previous equation with $\chi_{\delta}\geq 1$. To guarantee the existence of a such solution the following condition must be satisfied 
\begin{equation}
\gamma_\pm^2\leq\gamma^2\tanh\eta.
\end{equation}
Note that the previous condition holds only for $\gamma\geq\gamma_{\pm}$, and, since $\gamma_->\gamma_+$, if the condition with $\gamma_-$ holds the condition with $\gamma_+$ also does, but not conversely. 
Therefore, for $\gamma_+^2\leq\gamma^2\tanh\eta< \gamma_-^2$ we have only  extremal points $(u_+,u_-)=(\pm u_+^m,0)$, whereas for $\gamma_-^2\leq\gamma^2\tanh\eta$, in addition to the previous ones, a second pair of  extremal points appears at $(u_+,u_-)=(0,\pm u_-^m)$. The nature of these extremal points is unveiled for calculating the second derivatives.
From  Eq.(\ref{chi_delta}), if $u_{\mp}=0$ then 
\begin{equation}
\left(u_{\pm}^{m}\right)^{2}=\frac{\omega_{0}^{2}}{4\gamma^{2}(1\pm\delta)^{2}}\left(\chi_{\delta}^{2}-1\right)=\frac{\omega_{0}}{4\omega}\left(\frac{\gamma_{\pm}}{\gamma}\right)^{2}\left(\chi_{\delta}^{2}-1\right).
\end{equation}
Evaluating the second derivatives  at  these extremal points gives
\begin{eqnarray}
\left.\frac{\partial^{2} \phi_{\delta}}{\partial u_{\pm}^{2}}\right|_{(u_{\pm}^{m},0)}= +2\beta\omega\left(\frac{\chi_{\delta}^{2}-1}{\chi^{2}_{\delta}}\right)\left\{\eta\,\left[\left(\frac{\gamma}{\gamma_{\pm}}\right)^{2}-\left(\frac{\gamma_{\pm}}{\gamma}\right)^{2}\chi_{\delta}^{2}\right]-1\right\},\\ \nonumber
\left. \frac{\partial^{2} \phi_{\delta}}{\partial u_{\mp}^{2}}\right|_{(u_{\pm}^{m},0)}= -2\beta\omega\left[1 - \left(\frac{\gamma_{\pm}}{\gamma_{\mp}}\right)^{2}\right].\\ \nonumber
\left.\frac{\partial^{2} \phi_{\delta}}{\partial u_{+}\,\partial u_{-}} \right|_{(u_{\pm}^{m},0)}=0.
\end{eqnarray}
From these expressions  the Hessian's determinant is 
\begin{equation}
\fl D(u_{\pm}^{m},0)=-4\beta^{2}\omega^{2}\left[1 - \left(\frac{\gamma_{\pm}}{\gamma_{\mp}}\right)^{2}\right]\left(\frac{\chi_{\delta}^{2}-1}{\chi^{2}_{\delta}}\right)\left\{\eta\,\left[\left(\frac{\gamma}{\gamma_{\pm}}\right)^{2}-\left(\frac{\gamma_{\pm}}{\gamma}\right)^{2}\chi_{\delta}^{2}\right]-1\right\}.
\end{equation}
As $\gamma_{-}>\gamma_{+}$, for the  points $(\pm u_+^m,0)$ we have   $D(u_{+}^{m},0)>0$ with $\partial^{2}_{u_{-}}\phi_{\delta}<0$, whereas for   $(0,\pm u_-^m)$   $D(u_{-}^{m},0)<0$, then they correspond to  maximal and  saddle points, respectively. 
In the following we will deal with the maximal,
 $(u_+,u_-)=(\pm u_+^m,0)$, whose existence is guaranteed by the condition 
\begin{equation}
\gamma_+^2\leq\gamma^2\tanh\eta.
 \end{equation}
This condition defines the region in the parameters-temperature space corresponding to a first superradiant phase.
On the other hand, there is a pair of  saddle points that appears at $\gamma_-^2\leq\gamma^2\tanh\eta$. It defines a second superradiant  phase, which  coexists with the first one in the same parameters-temperature region. However, this second superradiant phase cannot be an equilibrium state from the thermodynamic point of view because it corresponds to a saddle-point. It does not have effects in the thermodynamics properties of the model, but its finite size corrections could make it detectable. The corresponding expressions for this second superradiant phase are very similar to the ones presented below for the thermally relevant superradiant phase. They  are shown in Appendix C.
 
\begin{figure}
\centering
\begin{tabular}{lll}
\includegraphics[width= 0.02\textwidth]{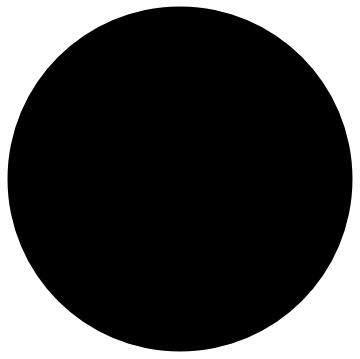}&\includegraphics[width= 0.02\textwidth]{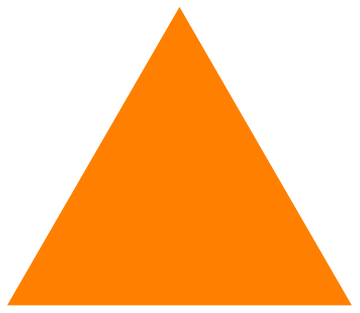}&\includegraphics[width= 0.02\textwidth]{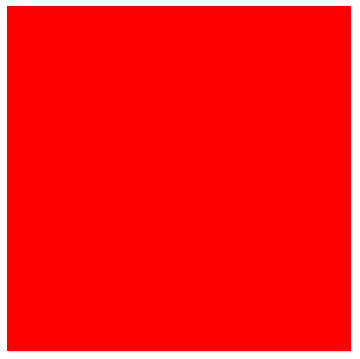}\\
\includegraphics[width= 0.32\textwidth]{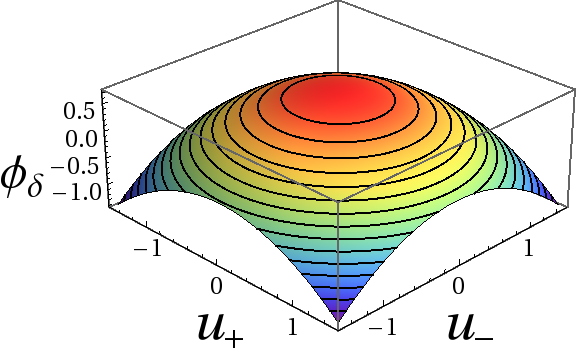}&\includegraphics[width= 0.3\textwidth]{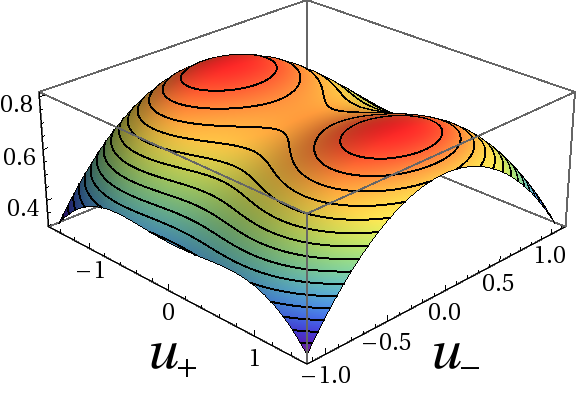}&\includegraphics[width= 0.3\textwidth]{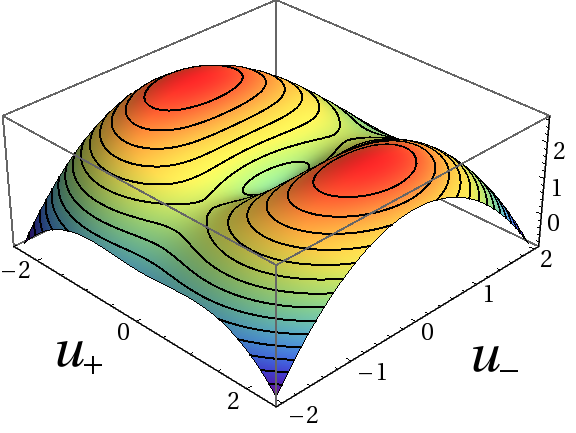}\\
\end{tabular} 
\includegraphics[width= 0.7\textwidth]{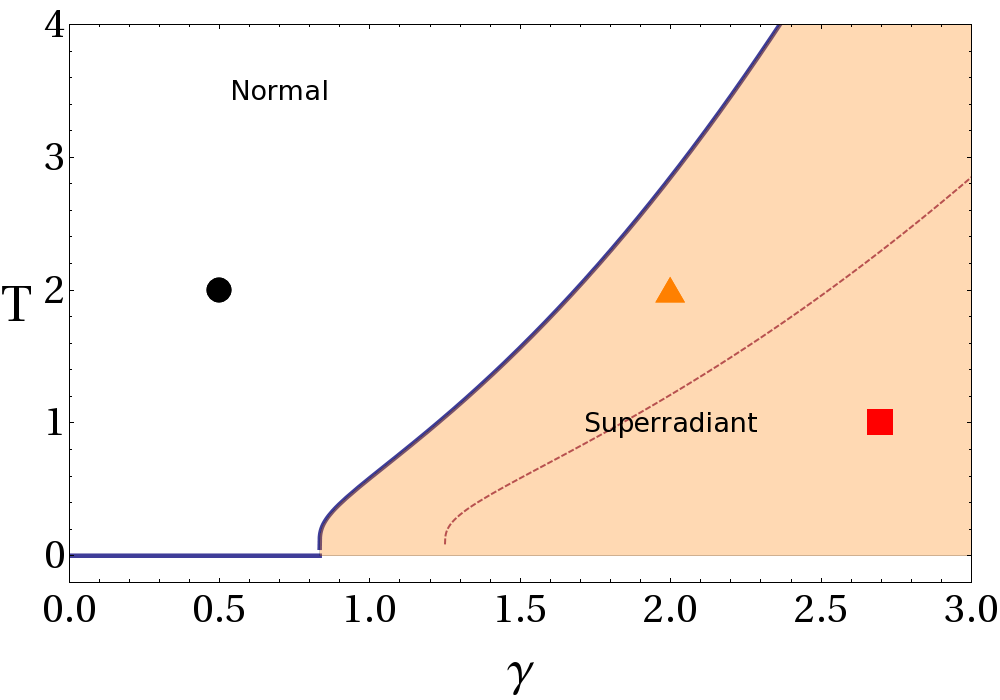}
\caption{Phases  in  temperature-coupling space (central panel) for $\delta=0.2$ and $\omega_0=\omega=1$, and 3D plots of function $\phi_\delta$ (top panels) for three representative points in phase space. Circle, triangle and square correspond, respectively, to $(\gamma,T)=(0.5,2),(2,2)$ and $(2.7,1)$. The solid curve in central panel, depicting the critical temperature, separates the normal phase, where the maximal point of function $\phi_\delta$  is located at $u_+^m=u_-^m=0$ (black circle in central  and   top panels gives a typical example), from the superradiant  region, where the maximal are located at $(u_+^m\not=0,u_-^m=0)$ (examples are indicated by the triangle and square). In the superradiant region, there exists another region (whose boundary is indicated by the red dashed line) where two saddle points of $\phi_\delta$, at $(u_+=0,u_-\not=0)$, emerge (red square gives an example).}
\label{fig:1}
\end{figure}

The transition between the normal and the first superradiant phase is given by the critical temperature defined in Eq.(\ref{betacrit}). In the region $\gamma_-^2\leq\gamma^2\tanh\eta$,  two saddle points at $(u_+,u_-)=(0,\pm u_-^m)$ appear,  but the maximal of $\phi_\delta$ are still given by  $(u_+,u_-)=(\pm u_+^m,0)$.  In figure  \ref{fig:1}, the normal and superradiant phases are shown in the $\gamma-T$ space for given values of the other parameters ($\delta, \omega$ and $\omega_o$). The surfaces  of the function $\phi_\delta$ are also plotted, where it can be seen that the maximal points change from $(u_+,u_-)=(0,0)$ in the normal phase to $(u_+,u_-)=(\pm u_+^m,0)$ in the superradiant one. The region where two saddle points appear is also indicated in the  diagram and illustrated by a representative $\phi_\delta$ surface.

It is important to emphasize what happens when we have the Tavis-Cummings case ($\delta=0$) and the Dicke case ($\delta=1$). This is reflected on the critical values of the coupling $\gamma_{\pm}$. In the first case, with $\delta=0$ we have $\gamma_{+}=\gamma_{-}$ then, only the normal and the superradiant phase with $\gamma_{0}=\sqrt{\omega_{0}\omega}$ exist. The integrable Tavis-Cummings case is the only one which has this feature. For every other value of $\delta$, which corresponds to non-integrable cases, there are two superradiant phases which can be distinguished. As $\delta$ tends to $1$ the critical value for the second superradiant goes to infinity, making this phase unobservable. Therefore, again we have only two phases, the normal and the superradiant phases, the latter marked, at $T=0$, by  $\gamma_{+}=\sqrt{\omega_{0}\omega}/2$.
 
The partition function in the first superradiant phase is calculated by the Laplace's integral method expanding the integrand around $(\pm u_+^m,0)$.  
By evaluating the function $\phi_\delta$ at  $(\pm u_+^m,0)$
$$
\phi_{\delta}(u_{+}^{m},0)= -\frac{\beta\omega_{0}}{4}\left(\frac{\gamma_{+}}{\gamma}\right)^{2}\left(\chi_{\delta}^{2}-1\right)+\ln\left[2\, \cosh\, \left(\eta\chi_{\delta}\right)\right],
$$  
we obtain the partition function after integration of the approximate gaussian function 
\begin{equation}
\fl \mathcal{Z}_{\delta}(T,\mathcal{N})=\mathcal{O}_{\delta,+}^{s.p.}(\mathcal{N})\,\exp\left[\mathcal{N}\left\{-\frac{\eta}{2}\left(\frac{\gamma_{+}}{\gamma}\right)^{2}\left(\chi_{\delta}^{2}-1\right)+\ln\left[2\, \cosh\, \left(\eta\chi_{\delta}\right)\right]\right\}\right],
\end{equation}
where 
\begin{eqnarray}
\mathcal{O}_{\delta,+}^{s.p.}(\mathcal{N})&=\frac{1}{\beta\omega}\left(1 - \left(\frac{\gamma_{+}}{\gamma_{-}}\right)^{2}\right)^{-1/2}\left(\frac{\chi_{\delta}^{2}-1}{\chi^{2}_{\delta}}\right)^{-1/2}\times\\ \nonumber
&\times \left\{1-\eta\,\left[\left(\frac{\gamma}{\gamma_{+}}\right)^{2}-\left(\frac{\gamma_{+}}{\gamma}\right)^{2}\chi_{\delta}^{2}\right]\right\}^{-1/2}_{\mbox{,}}
\end{eqnarray}
which gives a negligible contribution in the thermodynamic limit.
Then, the free energy is, 
\begin{equation}
-\beta\mathcal{F}(T)=-\frac{\beta\omega_{0}}{4}\left(\frac{\gamma_{+}}{\gamma}\right)^{2}\left(\chi_{\delta}^{2}-1\right)+\ln\left[2\, \cosh\, \left(\eta\chi_{\delta}\right)\right].
\end{equation}
We calculate the partition function's first derivative in the superradiant phase
\begin{equation}
\frac{1}{\beta\mathcal{Z}_{\delta}}\frac{\partial \mathcal{Z}_{\delta}}{\partial T}=\frac{1}{\beta\mathcal{O}_{\delta,+}^{s.p.}}\frac{\partial \mathcal{O}_{\delta,+}^{s.p.}}{\partial T}-k_{B}\mathcal{N}\,\frac{\eta}{2}\,\left(\frac{\gamma_{+}}{\gamma}\right)^{2}\left(\chi_{\delta}^{2}+1\right),
\end{equation}
to obtain  the entropy per particle
\begin{eqnarray}
\frac{\mathcal{S}_{\delta}(T)}{k_{B}}=-\eta\left(\frac{\gamma_{+}}{\gamma}\right)^{2}\chi_{\delta}^{2}+\ln\left[2\, \cosh\, \left(\eta\chi_{\delta}\right)\right]=\\
=-\eta\left(\frac{\gamma}{\gamma_{+}}\right)^{2}\tanh^{2}\left(\eta\,\chi_{\delta}\right)+\ln\left[2\, \cosh\, \left(\eta\chi_{\delta}\right)\right],
\end{eqnarray}
and from here, the energy per particle 
\begin{equation}
\fl \mathcal{U}_{\delta}(T)=-\frac{\omega_{0}}{4}\,\left(\frac{\gamma_{+}}{\gamma}\right)^{2}\left(\chi_{\delta}^{2}+1\right)=-\frac{\omega_{0}}{4}\,\left[\left(\frac{\gamma}{\gamma_{+}}\right)^{2}\tanh^{2}\left(\eta\,\chi_{\delta}\right)+\left(\frac{\gamma_{+}}{\gamma}\right)^{2}\right].
\end{equation}
By deriving implicitly (\ref{eq:sola}), it is straightforward to obtain the following expression for the heat capacity 
\begin{equation}
C_{\delta}=\frac{\partial \mathcal{U}_{\delta}}{\partial T}=k_{B}\eta^{2}\,\mbox{sech}^{2}\left(\eta\chi_{\delta}\right) \left(\frac{\gamma}{\gamma_{+}}\right)^{4}\frac{\tanh^{2}\left(\eta\,\chi_{\delta}\right)}{1-\left(\frac{\gamma}{\gamma_{+}}\right)^{2}\eta\,\mbox{sech}^{2}\left(\eta\,\chi_{\delta}\right)}.
\end{equation}

Regarding the thermal average of the photon number we note that the first integral, 
$u_-=0$, is different from zero only when $k=l$, so the average is 
\begin{eqnarray}
n_{\delta}^{k}=\sum_{l=0}^{k}{k \choose l}\delta_{k,l}\left(0\right)^{k-l}\left(u_{+}^{m}\right)^{l}=\left(u_{+}^{m}\right)^{k}=\\ \nonumber
=\left(\frac{\omega_{0}}{4\omega}\right)^{k}\left(\frac{\gamma_{+}}{\gamma}\right)^{2k}\left(\chi_{\delta}^{2}-1\right)^{k}=\\ \nonumber
n_{\delta}^{k}=\left(\frac{\omega_{0}}{4\omega}\right)^{k}\left(\left(\frac{\gamma}{\gamma_{+}}\right)^{2}\tanh^{2}\left(\eta\,\chi_{\delta}\right)-\left(\frac{\gamma_{+}}{\gamma}\right)^{2}\right)^{k}.
\end{eqnarray}

\begin{figure}
\centering
\begin{tabular}{ll}
(a) & (b) \\
\includegraphics[width=0.45\textwidth]{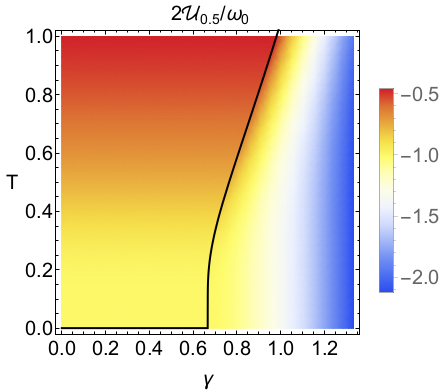}&\includegraphics[width=0.45\textwidth]{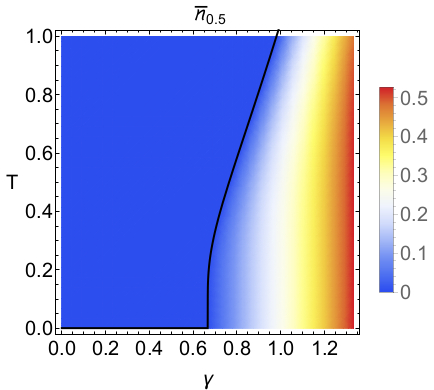} \\
(c) & (d) \\
\includegraphics[width=0.45\textwidth]{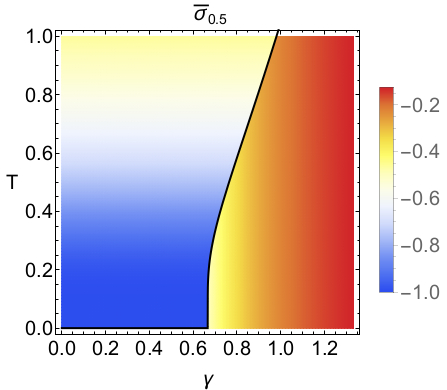}&\includegraphics[width=0.45\textwidth]{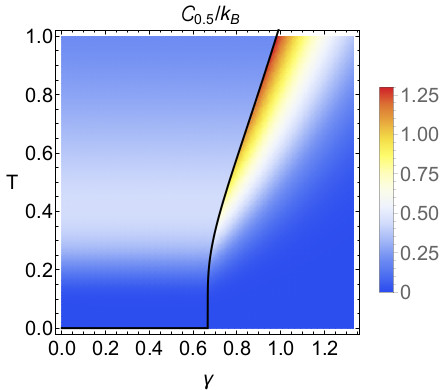} \\
\end{tabular} 
\caption{Density plots of the thermal averages per particle in the $(\gamma,T)$ plane. (a) internal energy, (b) photon number, (c) relative population, (d) heat capacity. For $\delta=0.5$, in resonance $\omega_{0}=\omega=1$. The thick black curve corresponds to the critical temperature which separates both the normal and the 
first
superradiant phase.}
\label{fig:2}
\end{figure}

Finally, for the collective atomic operators we have 
\begin{eqnarray}
\fl \sigma_{\delta,\mu}=-\frac{1}{2}\left(\frac{\gamma_{+}}{\gamma}\right)^{2}\left\{\delta_{\mu,z}+\frac{\gamma}{\omega_{0}}\left[(1+\delta)u_{+}\delta_{\mu,x}\right]\right\}=\\ \nonumber
\fl=-\frac{1}{2}\left(\frac{\gamma_{+}}{\gamma}\right)^{2}\left\{\delta_{\mu,z}+\frac{\gamma}{\omega_{0}}(1+\delta)\delta_{\mu,x}\sqrt{\frac{\omega_{0}}{4\omega}\left(\frac{\gamma_{+}}{\gamma}\right)^{2}\left(\chi_{\delta}^{2}-1\right)}\right\}=\\ \nonumber
\fl \sigma_{\delta,\mu}=-\frac{1}{2}\left(\frac{\gamma_{+}}{\gamma}\right)^{2}\left\{\delta_{\mu,z}+\frac{\gamma}{\omega_{0}}(1+\delta)\delta_{\mu,x}\sqrt{\frac{\omega_{0}}{4\omega}\left[\left(\frac{\gamma}{\gamma_{+}}\right)^{2}\tanh^{2}\left(\eta\,\chi_{\delta}\right)-\left(\frac{\gamma_{+}}{\gamma}\right)^{2}\right]}\right\}.
\end{eqnarray}
With these expressions and those corresponding to the normal phase, in Fig. \ref{fig:2} we show the density plots in the temperature vs. coupling space for the thermal averages of the most important observables, i. e., internal energy, photon number, relative population and heat capacity.

In order to write the entropy in terms of the energy, we note that 
\begin{equation}
\tanh\left(\eta\,\chi_{\delta}\right)=\left(\frac{\gamma_{+}}{\gamma}\right)^{2}\chi_{\delta}=\sqrt{-\left(\frac{\gamma_{+}}{\gamma}\right)^{2}\left(\frac{4\mathcal{U_{\delta}}}{\omega_{0}}+\left(\frac{\gamma_{+}}{\gamma}\right)^{2}\right)}.
\end{equation}
Then we have 
\begin{equation}
\eta=\frac{\mbox{arctanh}\left[\sqrt{-\left(\frac{\gamma_{+}}{\gamma}\right)^{2}\left(\frac{4\mathcal{U_{\delta}}}{\omega_{0}}+\left(\frac{\gamma_{+}}{\gamma}\right)^{2}\right)}\right]}{\left(\frac{\gamma}{\gamma_{+}}\right)^{2}\sqrt{-\left(\frac{\gamma_{+}}{\gamma}\right)^{2}\left(\frac{4\mathcal{U_{\delta}}}{\omega_{0}}+\left(\frac{\gamma_{+}}{\gamma}\right)^{2}\right)}}.
\end{equation}
So, the entropy is 
\begin{eqnarray}
\fl \frac{\mathcal{S}_{\delta}(\mathcal{U}_{\delta})}{k_{B}}=\ln(2)-\frac{1}{2}\ln\left\{1-\left[-\left(\frac{\gamma_{+}}{\gamma}\right)^{2}\left(\frac{4\,\mathcal{U}_{\delta}}{\omega_{0}}+\left(\frac{\gamma_{+}}{\gamma}\right)^{2}\right)\right]\right\}+\\ 
\fl -\mbox{arctanh}\left[\sqrt{-\left(\frac{\gamma_{+}}{\gamma}\right)^{2}\left(\frac{4\,\mathcal{U}_{\delta}}{\omega_{0}}+\left(\frac{\gamma_{+}}{\gamma}\right)^{2}\right)}\right]\sqrt{-\left(\frac{\gamma_{+}}{\gamma}\right)^{2}\left(\frac{4\,\mathcal{U}_{\delta}}{\omega_{0}}+\left(\frac{\gamma_{+}}{\gamma}\right)^{2}\right)}.
\end{eqnarray}
If we define
\begin{equation}
\mathcal{E}_{\delta}^{s.ph.+}=\sqrt{-\left(\frac{\gamma_{+}}{\gamma}\right)^{2}\left(\frac{4\,\mathcal{U}_{\delta}}{\omega_{0}}+\left(\frac{\gamma_{+}}{\gamma}\right)^{2}\right)},
\end{equation}
we have the same functional form for the entropy as in the normal phase 
\begin{equation}
\frac{\mathcal{S}(\mathcal{E}_{\delta}^{s.ph.+})}{k_{B}}=\ln(2)-\frac{1}{2}\ln\left[1-\left(\mathcal{E}_{\delta}^{s.ph.+}\right)^{2}\right]-\mathcal{E}_{\delta}^{s.ph.+}\,\mbox{arctanh}\left(\mathcal{E}_{\delta}^{s.ph.+}\right).
\end{equation}


\subsection{Phase Diagram in energy space                 }
\begin{figure}
\centering
\begin{tabular}{c}
\includegraphics[width=1.0\textwidth]{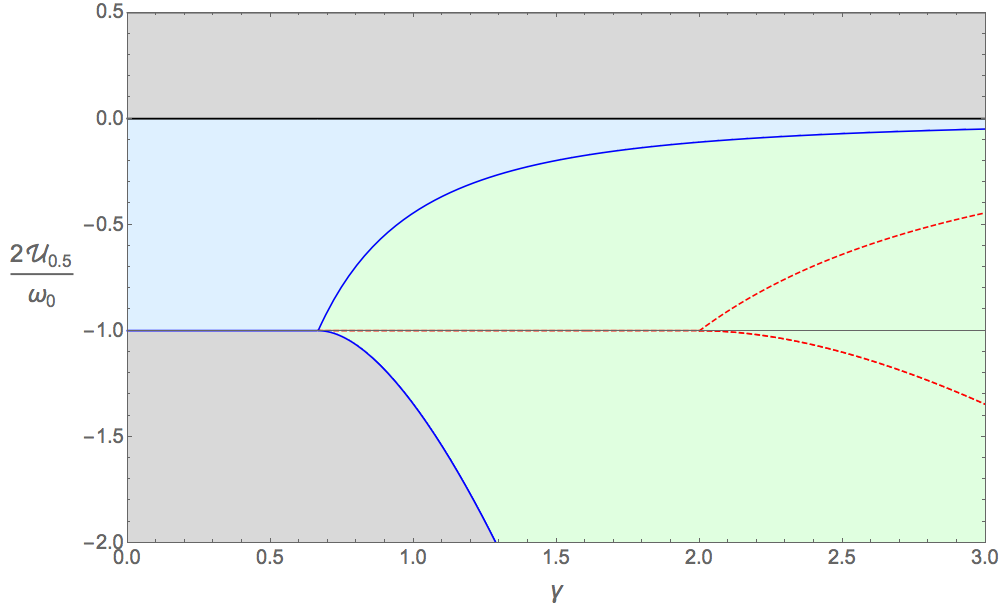} \\
\end{tabular} 
\caption{
Combined thermodynamical phases for the generalized Dicke model in the $\mathcal{U}_{\delta}$ vs. $\gamma$ space. We employ $\delta=0.5$. We can observe the normal phase (blue) and the first superradiant phase (green). The region where the second super radiant phase would be is marked through the red dotted lines. Not accesible zones are plotted as gray areas.}
\label{fig:3}
\end{figure}

First, we take the limits $T\rightarrow 0$ ($\beta\rightarrow\infty$) and $T\rightarrow \infty$ ($\beta\rightarrow 0$) over the expressions we have already found. In the limit $T\rightarrow 0$ the energy becomes
\begin{equation}
\label{inEn}
\frac{2\mathcal{U}_{\delta}(0)}{\omega_{0}}=\left\{
\begin{array}{cc}
-1 & \gamma < \gamma_{+}\\
-\frac{1}{2}\left[\left(\frac{\gamma}{\gamma_{+}}\right)^{2}+\left(\frac{\gamma_{+}}{\gamma}\right)^{2}\right] & \gamma\geq \gamma_{+}.
\end{array}
\right. .
\end{equation}
We note that, for $\delta=0,1$, we recover in a straightforward way the well-known Quantum Phase Transition (QPT) of the Dicke and Tavis-Cummings models. At the same limit, the thermal averages of the photon number and pseudo-spin collective operators are 
\begin{eqnarray}
\fl 2\sigma_{\delta,\mu}(0)&=\left\{
\begin{array}{cc}
-\delta_{\mu,z} & \gamma < \gamma_{+}\\
-\left(\frac{\gamma_{+}}{\gamma}\right)^{2}\left(\delta_{\mu,z}+\frac{\gamma}{\omega_{0}}(1-\delta)\delta_{\mu,x}\,\sqrt{\frac{\omega_{0}}{4\omega}\left[\left(\frac{\gamma}{\gamma_{+}}\right)^{2}-\left(\frac{\gamma_{+}}{\gamma}\right)^{2}\right]}\right)& \gamma\geq \gamma_{+}\\
\end{array}
\right. ,\\
\fl n_{\delta}^{k}(0)&=\left\{
\begin{array}{cc}
0 & \gamma < \gamma_{+}\\
\left(\frac{\omega_{0}}{4\omega}\right)^{k}\left[\left(\frac{\gamma}{\gamma_{+}}\right)^{2}-\left(\frac{\gamma_{+}}{\gamma}\right)^{2}\right] & \gamma\geq \gamma_{+}
\end{array}
\right. .
\end{eqnarray}

On the other hand, in the limit of infinite temperature we have 
\begin{eqnarray}
\lim_{T\rightarrow\infty}\frac{2\mathcal{U}_{\delta}}{\omega_{0}}&=0,\\
\lim_{T\rightarrow\infty}2\sigma_{\delta}(0)&=0,\\
\lim_{T\rightarrow\infty} n_{\delta}^{k}(0)&=0.
\end{eqnarray}

In this limit, the atomic space is saturated and  the energy has an upper limit $\mathcal{U}_{\delta}=0$. 

Finally, we note that, at the critical temperature, the energy takes the value
\begin{equation} \label{eqn:ec}
\frac{2\mathcal{U}_{\delta,c}}{\omega_{0}}=\frac{2\mathcal{U}_{\delta}(T_{c})}{\omega_{0}}=-\left(\frac{\gamma_{+}}{\gamma}\right)^{2}.
\end{equation}
Then, in the space of $\mathcal{U}_{\delta}$ we have a critical energy as a function of $\gamma$ which separates the normal from the superradiant phase.
 
 In Fig. \ref{fig:3} we show the phase diagram in the $\mathcal{U}_{\delta}$ vs.$\gamma$ space. The region where the saddle points $(u_+,u_-)=(0,\pm u_-^m)$ of function $\phi_\delta$ appear is also indicated.  Even if the emergence of these saddle points has no consequences in  the thermodynamics of the model, they can be linked to a singular behavior of the models's  density of states as it will be discussed in the following section.    

\begin{figure}
\centering
\begin{tabular}{c}
\includegraphics[width=0.7\textwidth]{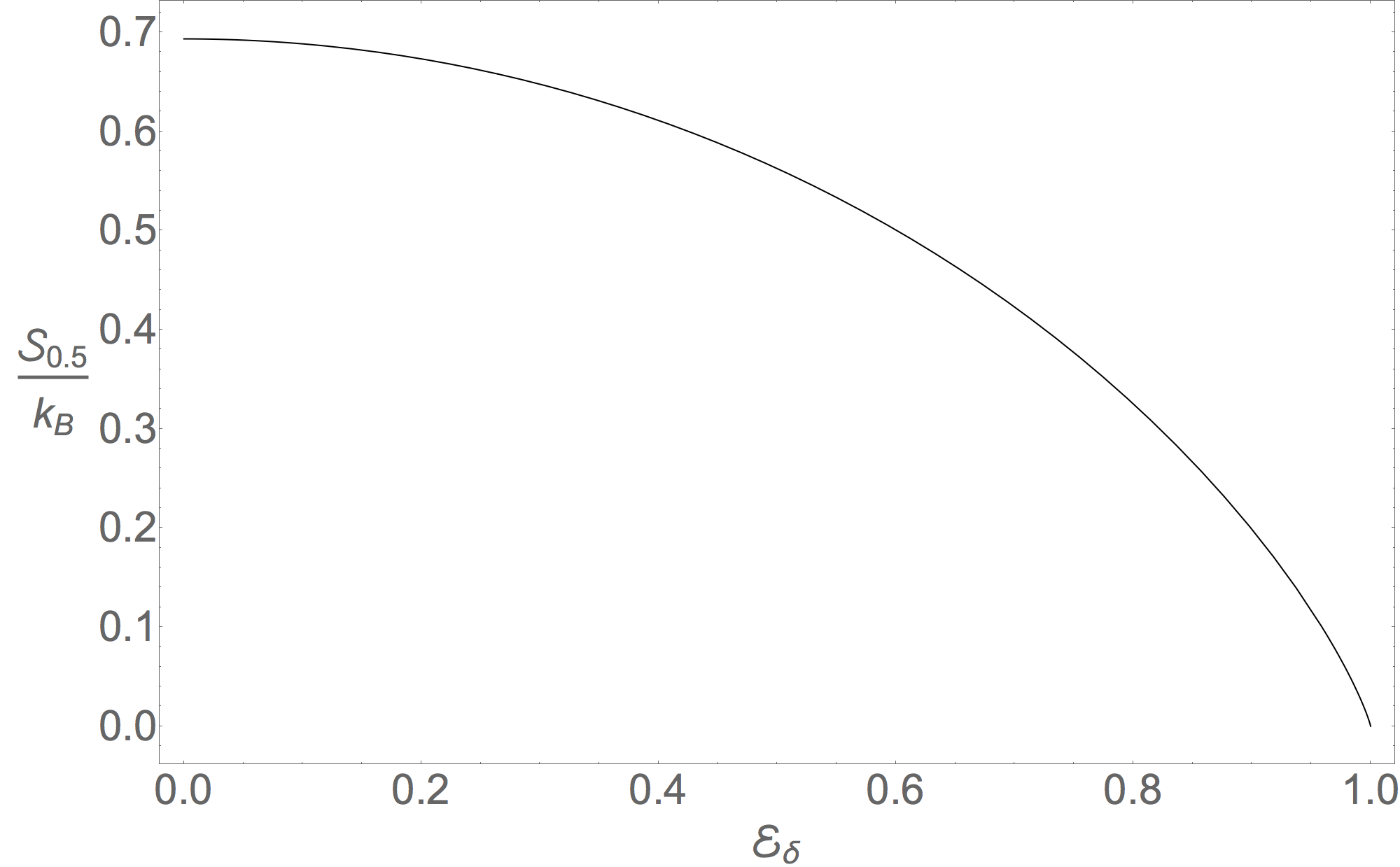}
\end{tabular} 
\caption{Entropy per particle $\mathcal{S}_{\delta}$ as a function of scaled energy variable $\mathcal{E}_{\delta}$. The same functional form occurs for both the normal and first superradiant phases.}
\label{fig:4}
\end{figure}

Before closing this section, let us discuss an expression for the entropy  as a function of the energy. Above it was shown that the entropy can be put at the same form  regardless of the phase   
\begin{eqnarray}\label{entrocan}
\frac{\mathcal{S}(u_{\delta})}{k_{B}}&=&\ln(2)-\frac{1}{2}\ln\left[1-\left(\mathcal{E}_{\delta}\right)^{2}\right]-\mathcal{E}_{\delta}\,\mbox{arctanh}\left(\mathcal{E}_{\delta}\right)\nonumber\\
 &=& \ln(2)-\frac{1}{2}\left(1+\mathcal{E}_{\delta}\right)\,\ln\left(1+\mathcal{E}_{\delta}\right)-\frac{1}{2}\left(1-\mathcal{E}_{\delta}\right)\,\ln\left(1-\mathcal{E}_{\delta}\right).
\end{eqnarray}
where we have used the identity $\mbox{arctanh}(x)=\frac{1}{2}\,\ln\left(\frac{1+x}{1-x}\right)$, and the function  $\mathcal{E}_\delta$ is given by 

\begin{equation}\label{funu}
\fl \mathcal{E}_{\delta}=\left\{\begin{array}{cc}
-\frac{2\mathcal{U}_{\delta}}{\omega_{0}} &  {\hbox {for the normal phase }} \\
& (\gamma<\gamma_{+}\,\,\,\mbox{or}\,\,\,\gamma\geq\gamma_{+}\,\,\mbox{with}\,\, \mathcal{U}_{\delta}>\mathcal{U}_{\delta,c}) \\
\sqrt{-\left(\frac{\gamma_{+}}{\gamma}\right)^{2}\left(\frac{4\,\mathcal{U}_{\delta}}{\omega_{0}}+\left(\frac{\gamma_{+}}{\gamma}\right)^{2}\right)} &  {\hbox {for the superradiant phase }} \\
& (\gamma\geq\gamma_{+}\,\,\mbox{with }\,\, \mathcal{U}_{\delta}\leq\mathcal{U}_{\delta,c})
\end{array}\right. 
\end{equation}

In Fig. \ref{fig:4} it is shown the functional behavior of the entropy per particle as a function of the scaled energy variable $\mathcal{E}_{\delta}$. 
We will reproduce this result and  give a simple meaning to the function $\mathcal{E}_{\delta}$ in the next section, where  the thermodynamics of the model is obtained  from  a microcanonical approach.

\section{Micro-canonical thermodynamics}

In the microcanonical ensemble, a complete thermodynamic representation is given in terms of the entropy 
\begin{equation} \label{ent}
S(E,\mathcal{N})=k_{B}\,\ln\left[\frac{}{}\Omega(E,\mathcal{N})\right],
\end{equation}
where $\Omega(E,\mathcal{N})$ is the number of states for a given energy $E$ and number of atoms $\mathcal{N}$. $k_{B}$ is the Boltzmann constant. For a system of $\mathcal{N}$ distinguishable two-level atoms we have $2^{\mathcal{N}}$ states distributed over all the subspaces of total pseudo-spin identified through $j$. Each subspace, $j$, has a number of states given by the multiplicity $Y(\mathcal{N},j)$. We calculate $\Omega(E,\mathcal{N})$ through the following formula, 
\begin{equation}
\Omega(E,\mathcal{N})=\sum_{j=0}^{\mathcal{N}/2}Y(\mathcal{N},j)\,\nu(E,\mathcal{N},j) \Delta E,
\end{equation}
where $\nu(E,\mathcal{N},j) \Delta E $ is the number of states in  the  energy interval $[E,E+\Delta E]$,  for the pseudo-spin $j$. For each $j$ we approximate $\nu(E,\mathcal{N},j)$ by means of the semi-classical Density of States (SDoS) obtained by integration of the available phase-space volume, which is the semiclassical leading order of  the Gutzwiller-trace formula \cite{Gutz90,Bas14,Str15}.  

In order to calculate the thermodynamics of the generalized Dicke model, in addition to the density of states (DoS) for  given $j$ and energy, $\nu(E,\mathcal{N},j)$, we have to estimate the multiplicity $Y(\mathcal{N},j)$ which gives the number of different ways that a set of  $\mathcal{N}$ spin $1/2$ systems can couple to a total pseudospin $j$.  We focus, first,  on the former  quantity.  Even if, as we will demonstrate,  the thermodynamics of the model is entirely dominated by the multiplicity $Y(\mathcal{N},j)$ and the dependence on $\nu(E,\mathcal{N},j)$ is completely diluted in the thermodynamical limit,  for the sake of completeness  and for  future reference for finite size studies, we will give entire   expression for the DoS using a semiclassical approximation. 

It can be seen (see Appendix D) that the quantum density of states 
\begin{equation} \label{a}
\nu_Q \left(E,\mathcal{N},j\right) = \sum_{n} \delta \left(E - E^{j}_{n}\right), 
\end{equation}
with $ E^{j}_{n}$ the quantum spectrum for a given $j$, can be semiclassically  approximated ($\hbar=1$) by
 \begin{equation}\label{s}
\nu_\delta \left(E, \mathcal{N},j\right) = \frac{1}{\left(2\pi\right)^{2}} \int \delta\left[E - H_{cl;j}\left(z,\alpha\right)\right] dq_-dq_+~dj_{z}~d\phi.
\end{equation}
This expresion defines the semiclassical density of states (SDoS) we study here.
In the SDoS appear $H_{cl,j}\left(z,\alpha\right)$, which  is the expectation value of the generalized Dicke Hamiltonian in  Glauber ($|\alpha\rangle$) and Bloch ($|z,j\rangle$) coherent states for the photonic and atomic parts respectively \cite{Bas14}
\begin{eqnarray}
\fl & H_{cl,j}\left(z,\alpha\right)=\langle \alpha z,j | H_{D,\delta}|\alpha z,j\rangle= \\
\fl &=\omega|\alpha|^{2} -\omega_{0} j\left(\frac{1-|z|^{2}}{1+|z|^{2}}\right) + \frac{\gamma}{\sqrt{N}}\left[ (1+\delta)(\alpha+\alpha^*)\frac{Re(z)}{1+|z|^{2}}-i(1-\delta)(\alpha-\alpha^*)\frac{Im(z)}{1+|z|^{2}} \right].\nonumber
\end{eqnarray}
The variables $q_\pm$ and $j_z$,$\phi$ are canonical variables related with the Glauber and Bloch coherent parameters through
\begin{equation}
\label{paraCoh}
\alpha=\frac{1}{\sqrt{2}}\left(q_{+}+iq_{-}\right),\  \ \ \ z=\sqrt{\frac{1+\frac{j_{z}}{j}}{1-\frac{j_{z}}{j}}}e^{-i\phi}.
\end{equation}
The variables $q_\pm$ can take any real value, whereas the Bloch variables are restricted to the intervals $\phi\in[0,2 \pi)$ and $j_z\in[-j,j]$.
The Hamiltonian written in these variables is
\begin{eqnarray} \label{Hclass}
H_{cl,j}&=\omega_{0}j_{z}+\frac{\omega}{2}\left(q_{+}^{2}+q_{-}^{2}\right)+\frac{\gamma\,j}{\sqrt{\mathcal{N}/2}}\sqrt{1-\left(\frac{j_{z}}{j}\right)^{2}}\times\\ \nonumber
&\times \left[(1+\delta)q_{+}\,\cos(\phi)-(1-\delta)q_{-}\,\sin(\phi)\right].
\end{eqnarray}

\subsection{Lowest energies for each $j$}

The previous Hamiltonian is clearly only lower bounded $H_{cl,j}\in [E^{gs}_j(\gamma, \delta),\infty)$. To obtain the semiclassical lowest energy $E^{gs}_j(\gamma,\delta)$ for each $j$,  we calculate its derivatives  and make $\nabla H_{cl}(q_{-}^{m},q_{+}^{m},j_{z}^{m},\phi^{m})=0$. By solving this set of equations, we obtain the semiclassical lowest energies (for details see Appendix E)
\begin{eqnarray} \label{eqn:gse}
E_{j}^{gs}(\gamma,\delta)=\left\{
\begin{array}{cc}
-j\omega_{0} & \gamma<\gamma_{j,+} \\
E_{j,e}^+ & \gamma\geq\gamma_{j,+}\\
\end{array}
\right. ,
\end{eqnarray} 
 
where we have used the definitions 
\begin{equation}\label{defgamm}
\fl \gamma_{j,\pm}\equiv\sqrt{\frac{\mathcal{N}}{2j}}\frac{\sqrt{\omega_{0}\omega}}{\left(1\pm\delta\right)}=\sqrt{\frac{\mathcal{N}}{2j}}\,\gamma_{\pm},
\mbox{  \ \ \    and   \ \ \ \ }
E_{j,e}^{\pm}\equiv-\frac{j\omega_{0}}{2}\left[\left(\frac{\gamma_{j,\pm}}{\gamma}\right)^{2}+\left(\frac{\gamma}{\gamma_{j,\pm}}\right)^{2}\right].
\end{equation}
The corresponding values of the canonical variables that minimize the energy are given by
\begin{equation}\label{point:gse}
\fl (q_{+}^{min},q_{-}^{min},j_{z}^{min},\phi^{min})=\left\{\begin{array}{cc} (0,0,-j,\mbox{arbitrary}) &\gamma<\gamma_{j,+}\\
\left(\pm q_+^m,0,-j\left(\frac{\gamma_{j,+}}{\gamma}\right)^{2},0\,\,\mbox{or}\,\,\pi\right)& \gamma\geq\gamma_{j,+}
\end{array}\right. ,
\end{equation}
where $q_+^m$ is defined by
\begin{equation}
q_\pm^m=\frac{\gamma (1\pm\delta)}{\omega}\frac{j}{\sqrt{\mathcal{N}/2}}\sqrt{1-\left(\frac{\gamma_{j,\pm}}{\gamma}\right)^{4}}.
\end{equation}

From the coordinate values that minimize the energy (\ref{point:gse}), it is clear that the lowest energy states for  $\gamma<\gamma_{j,+}$ correspond to  states with zero photons [$\langle\alpha|a^\dagger a|\alpha\rangle=(q_+^2+q_-^2)/2=0$], whereas the lowest energy states for the cases $\gamma>\gamma_{j,+}$ have a mean number of photons different to zero given by $\langle\alpha|a^\dagger a|\alpha\rangle=(q_+^m)^2/2$. Therefore, $\gamma_{j,+}$ is a critical coupling that separates, for  a given $j$, the  normal and superradiant phases.

\begin{figure}
\centering
\begin{tabular}{c}
\includegraphics[width=0.6\textwidth]{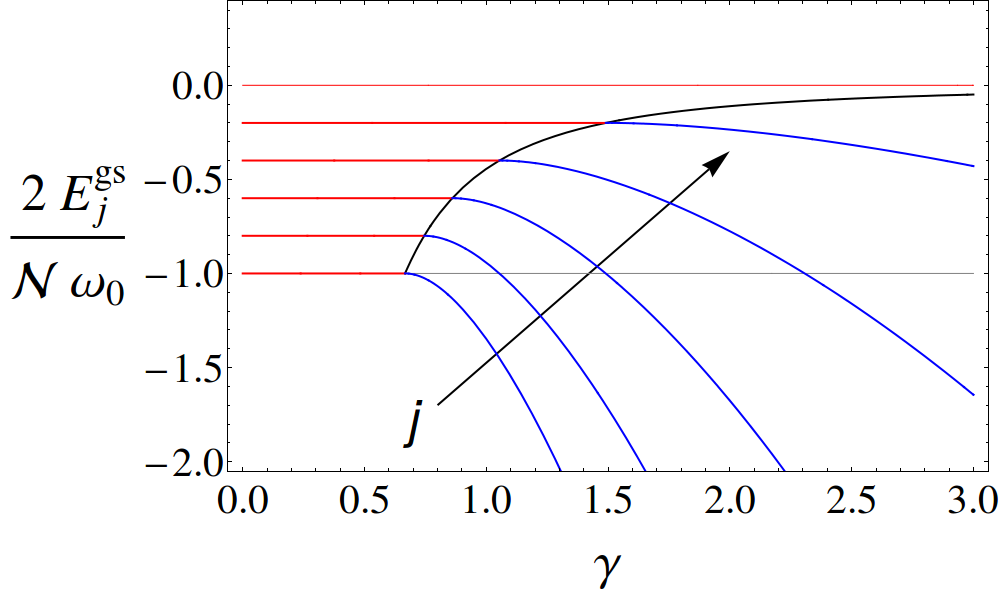}\\\includegraphics[width=0.6\textwidth]{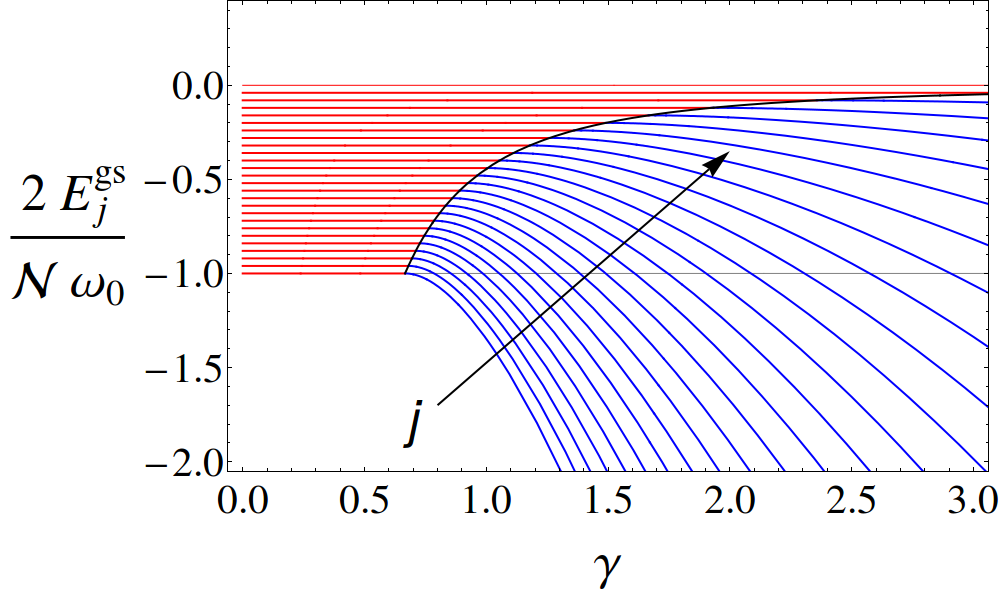}
\end{tabular} 
\caption{Lowest energies for each pseudospin $j$ as a function of the coupling $\gamma$, for $\mathcal{N}=10$ (left) and $\mathcal{N}=50$ (right) with $\delta=0.5$. Each curve is colored in red in the normal phase ($\gamma<\gamma_{j,+}$) and in blue for the superradiant one ($\gamma_{j,+}\leq \gamma$). The line separating these  two phases, Eq.(\ref{eqpt}), is plotted in black.  The arrow indicates the descending order of $j=\mathcal{N}/2,...,2,1,0$. }
\label{fig:5}
\end{figure}

In Fig.\ref{fig:5} we plot the semiclassical lowest energy for every $j$ ($j=0,1,...,\mathcal{N}/2$) for two different number of atoms. Every curve is colored in red when $\gamma<\gamma_{j,+}$ and in blue for $\gamma>\gamma_{j,+}$. Observe that, for a given coupling, the minimal energy increases as $j$ decreases, making the maximal pseudo-spin $j=\mathcal{N}/2$, the global ground state

\begin{eqnarray} \label{eqn:gsGlobal}
E_{min}(\gamma,\delta)=E_{j=\mathcal{N}/2}^{gs}(\gamma,\delta)=\left\{
\begin{array}{cc}
-\mathcal{N}\omega_{0}/2 & \gamma<\gamma_{+} \\
-\frac{\mathcal{N}\omega_{0}}{4}\left[\left(\frac{\gamma_{+}}{\gamma}\right)^{2}+\left(\frac{\gamma}{\gamma_{+}}\right)^{2}\right] & \gamma\geq\gamma_{+}.\\
\end{array}
\right. 
\end{eqnarray} 

It is interesting to find the other extremal points of the semiclassical Hamiltonian $H_{cl,j}$ (see Appendix E) because, as we will see below, they signal  the energy values where a singular behaviour of the density of states is observed. The complete classification of these points is given in Table 1, where we can identify, for every $j$ three different intervals for the coupling $\gamma$, {\bf a)} $\gamma<\gamma_{j,+}$, {\bf b)} $\gamma_{j,+}\leq\gamma<\gamma_{j,-}$ and {\bf c)} $\gamma_{j,-}\leq \gamma$, where  
$\gamma_{j,-}$, defined above (\ref{defgamm}), clearly satifies $\gamma_{j.-}\geq\gamma_{j,+}$.
\begin{table}
\footnotesize
\begin{tabular}{|c|l|c|c|}
\hline
 interval& extremal points $(q_+,q_-,j_z,\phi)$ & Energy $E$ & Type \\
 \hline
 \hline
$\gamma<\gamma_{j,+}$& {\bf{1)}} $(0,0,-j,\mbox{arbitrary})$  & $-j\omega_o$ & global minimum (ground state) \\
                     & {\bf{2)}} $(0,0,+j,\mbox{arbitrary})$  & $+j\omega_o$ & local maximum\\
                     \hline 
$\gamma_{j,+}\leq\gamma <\gamma_{j,-}$ & {\bf{1)}} $(\mp q_+^ m,0,-j\left(\frac{\gamma_{j,+}}{\gamma}\right)^2,0 \mbox{ or } \pi)$  & $E_{j,e}^+$ & global minima (ground state) \\
                                       & {\bf{2)}} $(0,0,-j,\mbox{arbitrary}) $ &$-j\omega_o$ & saddle point\\
                                       & {\bf{3)}} $(0,0,+j,\mbox{arbitrary}) $ & $j\omega_o$& local maximum\\
\hline
$\gamma_{j,-}\leq\gamma $ & {\bf{1)}} $(\mp q_+^ m,0,-j\left(\frac{\gamma_{j,+}}{\gamma}\right)^2,0 \mbox{ or } \pi)$  & $E_{j,e}^+$  & global minima (ground state) \\
                          & {\bf{2)}} $(0,\pm q_-^m, -j\left(\frac{\gamma_{j,-}}{\gamma}\right)^2, \pi/2\mbox{ or }3\pi/2)$ &$E_{j,e}^-$  & saddle points\\ 
                          & {\bf{3)}} $ (0,0,-j,\mbox{arbitrary})$ &$-j\omega_o$ & local maximum\\
                          & {\bf{4)}} $ (0,0,+j,\mbox{arbitrary})$ & $j\omega_o$& local maximum\\
                          \hline\hline
\end{tabular}
\caption{Classification of the extremal points of semiclassical Hamiltonian $H_{cl,j}$ for the different intervals in the parameter $\gamma$. The arbitrariness of variable $\phi$ in several extremal points comes from the fact that these points are the north and south poles of the Bloch sphere, where the value of the azimuthal angle $\phi$ is completely irrelevant.}
\end{table}
\normalsize


\subsection{Quantum and thermal phase transtions}
Before calculating the density of states, some interesting preliminary relations between the previous results for the lowest energies at each $j$ and the thermodynamics of the model, can be established. First, for a given coupling, the global minimal energy  (\ref{eqn:gsGlobal}) is equal to the internal energy at $T=0$ (\ref{inEn}) found in the canonical ensemble calculation of the previous section. For the other pseudospins,  similar QPTs are observed, but at larger coupling values, which are given by $\gamma=\gamma_{j,+}=\sqrt{\frac{\mathcal{N}}{2 j}}\gamma_+$ (\ref{defgamm}). The energies where these QPTs occur are given by $E=-j \omega_0$ (\ref{eqn:gse}). Combining these latter two  expressions, we obtain a curve in the space $E$ vs. $\gamma$, where the $QPT$s for the ground-states occur  for every pseudospin $j$  
\begin{equation}\label{eqpt}
\frac{2E_{j}^{QPT}}{\omega_{0}\mathcal{N}}=-\left(\frac{\gamma_{+}}{\gamma}\right)^{2}.
\end{equation}
This curve is the same we found before (\ref{eqn:ec}), in the canonical thermodynamical approach,  for the internal energy evaluated at the  critical  temperature. 

\begin{figure}
\centering
\begin{tabular}{c}
\includegraphics[width=1.0\textwidth]{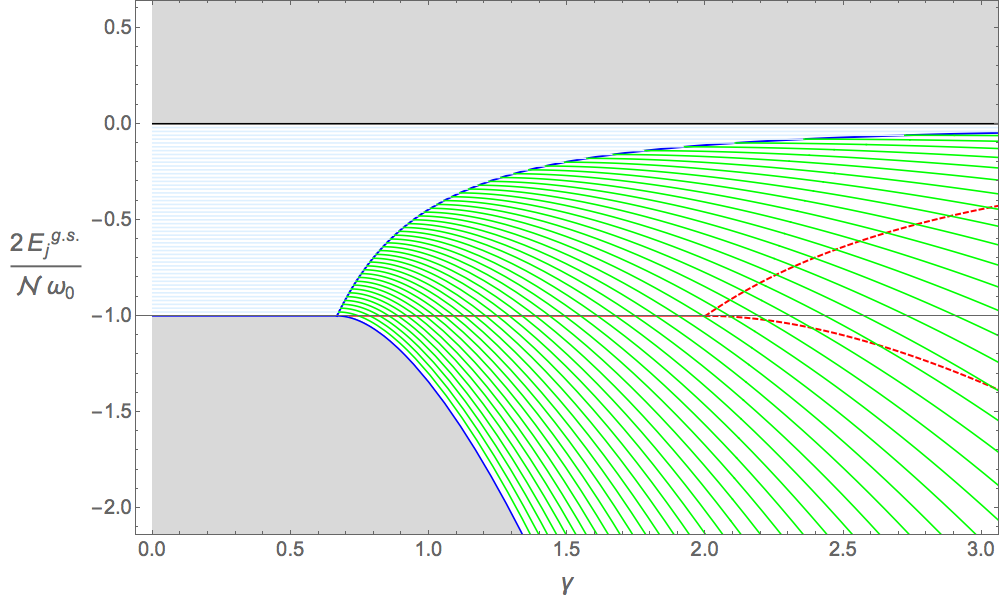} \\
\end{tabular} 
\caption{Phase diagram showing the lowest
energies of each  $j$ in the energy vs. coupling space. It reproduces the thermodynamical phase space of Fig. \ref{fig:3}.}
\label{fig:6}
\end{figure}

Therefore, the critical energy which separates the different phases in the thermodynamical space is formed by the aggregation of the individual $QPT$s of each pseudospin $j$. We can conclude that the thermodynamic meaning of the  $QPT$s is the thermal superradiant phase transition, when all the pseudo spin subspaces are properly included in the analysis. This relation  between the $QPT$s and the thermal phase transitions can be visualized in Fig. \ref{fig:6},  where we reproduce the thermodynamical phase diagram using only the information of the lowest energies for each pseudo-spin sector $j$.  

In the following sections, the previous observations will be put in more solid grounds.  In order to that, a simple observation is important. Note that   for a given energy $E<0$ not all the pseudo-spins are available, only the largest pseudos-spins that satisfies $E\geq E_j^{gs}$ can participate at that given energy. In the case  $E>0$,  the previous condition is satisfied for every $j$, making this energy region, as it will be shown below, thermodynamically  inaccessible in accord with the cannonical ensemble result of Fig.\ref{fig:3}.    
 
\subsection{Semi-classical Density of States}

Following the Appendix A of reference \cite{Bas14}, the $q_\pm$ integral of the   SDoS (\ref{s}) can be easily performed to obtain  for the generalized Dicke model, the following  expression up to the integration of the atomic classical variables 
\begin{equation}
\nu_\delta(E,\mathcal{N},j)=\frac{1}{2\pi\,\omega}\int\,dj_{z}\int\,d\phi.
\end{equation}

The dependence on $E$ of the previous integral, comes from the bounds of the atomic classical variables $j_z$ and $\phi$. A detailed analysis (see Appendix F) allows to determine these bounds for the different regimes in coupling and energy space. The different energy regimes are defined by the extremal energies of the semiclassical Hamiltonian $H_{cl,j}$   shown in Table 1. In reference \cite{Bran13} equivalent expressions to the SDoS, shown below, were obtained, using an inverse Laplace transformation to the partition function to obtain them. 

\begin{figure}
\centering
\begin{tabular}{cc}
a) & b)\\
\includegraphics[width=.4\textwidth]{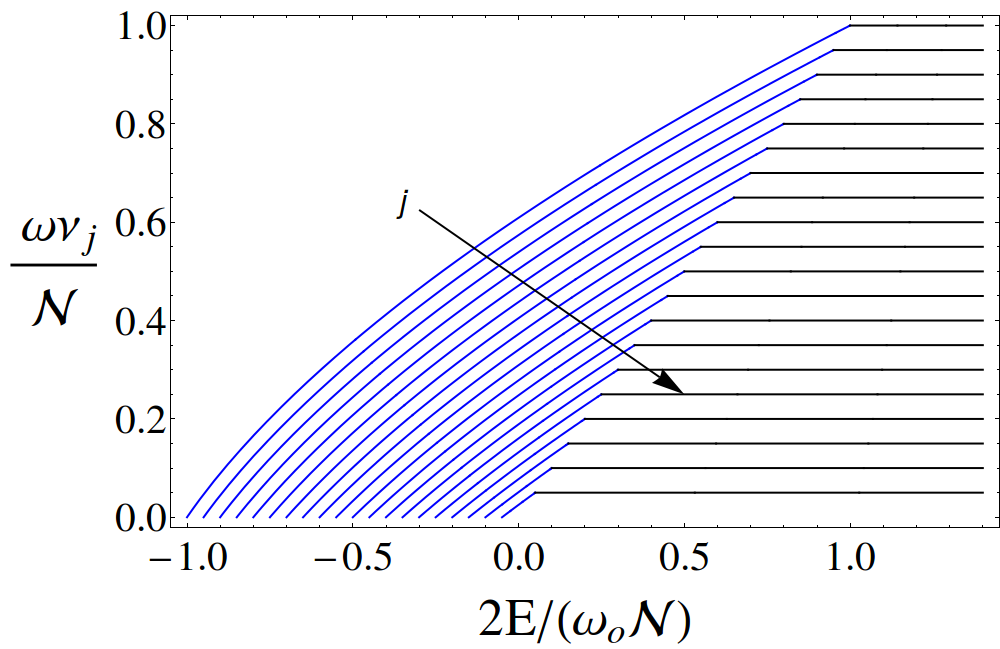}& \includegraphics[width=.4\textwidth]{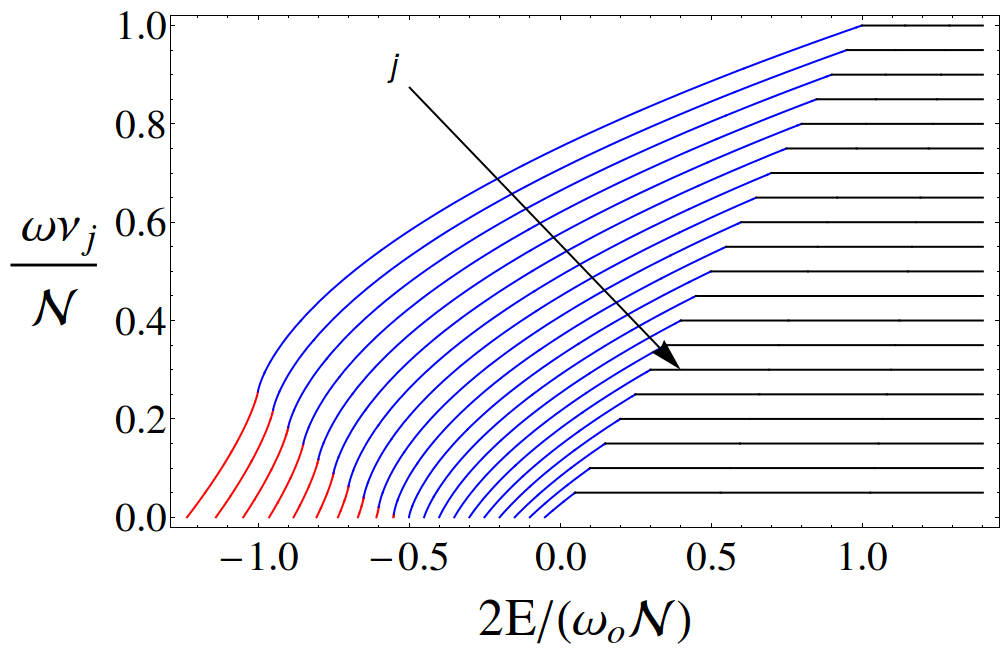}\\
c)& d)\\
\multirow{3}{*}{\includegraphics[width=.4\textwidth]{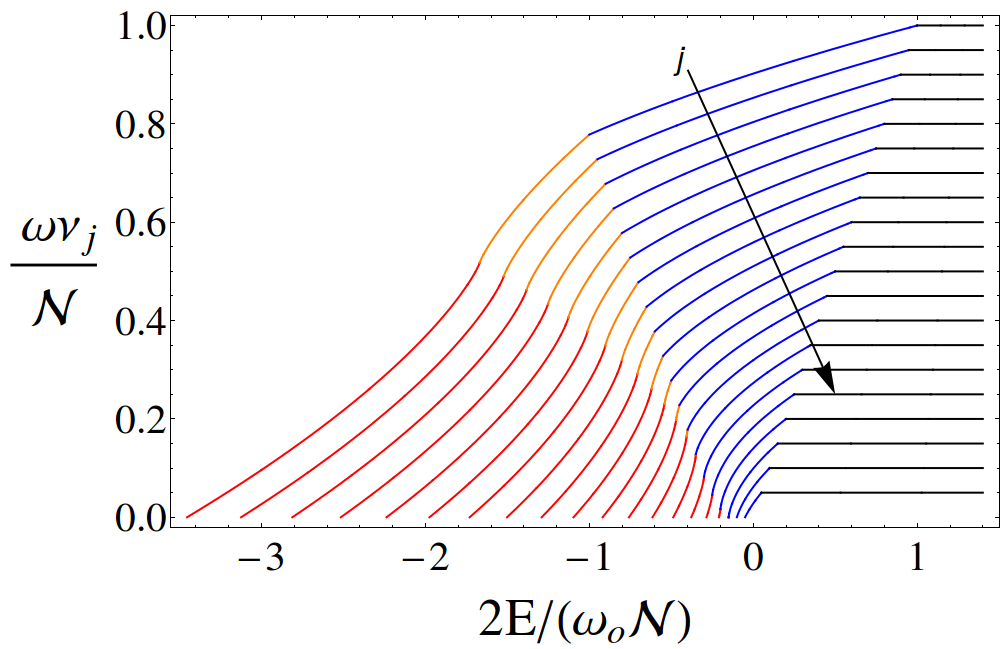}} & \\
&\includegraphics[width=.17\textwidth]{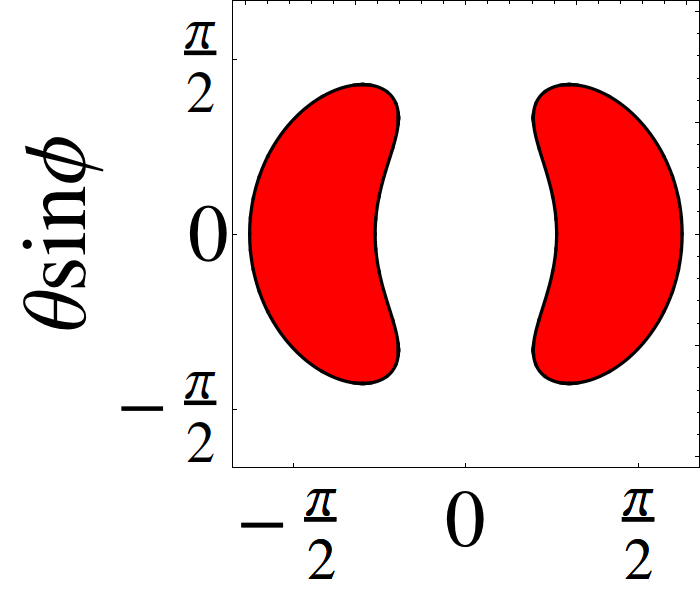}\includegraphics[width=.145\textwidth]{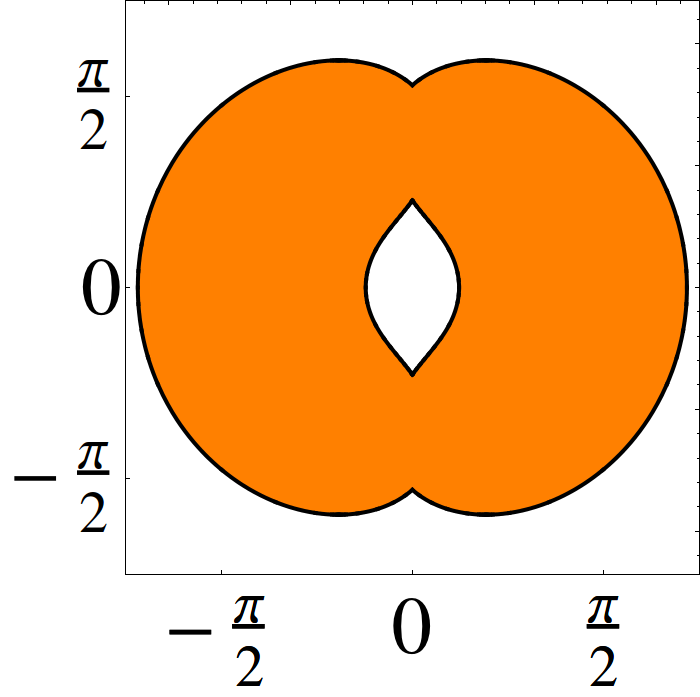}\\
&  \includegraphics[width=.16\textwidth]{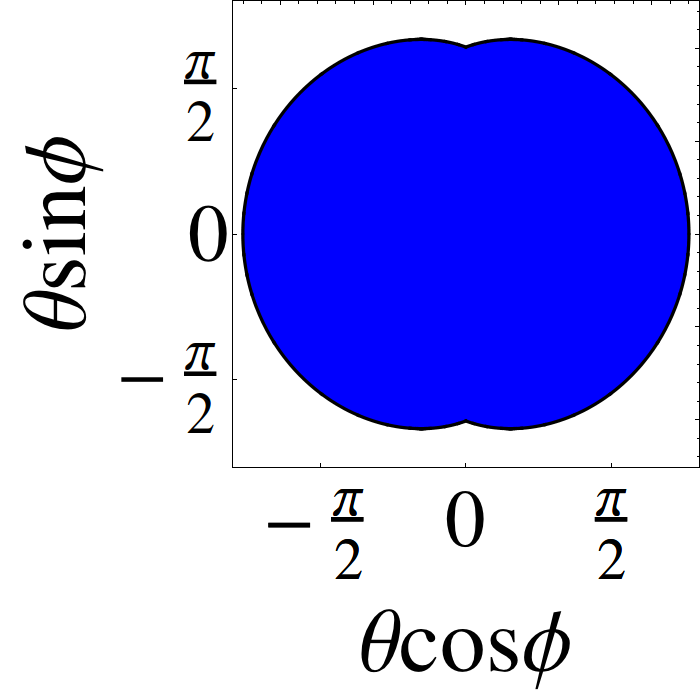}\includegraphics[width=.14\textwidth]{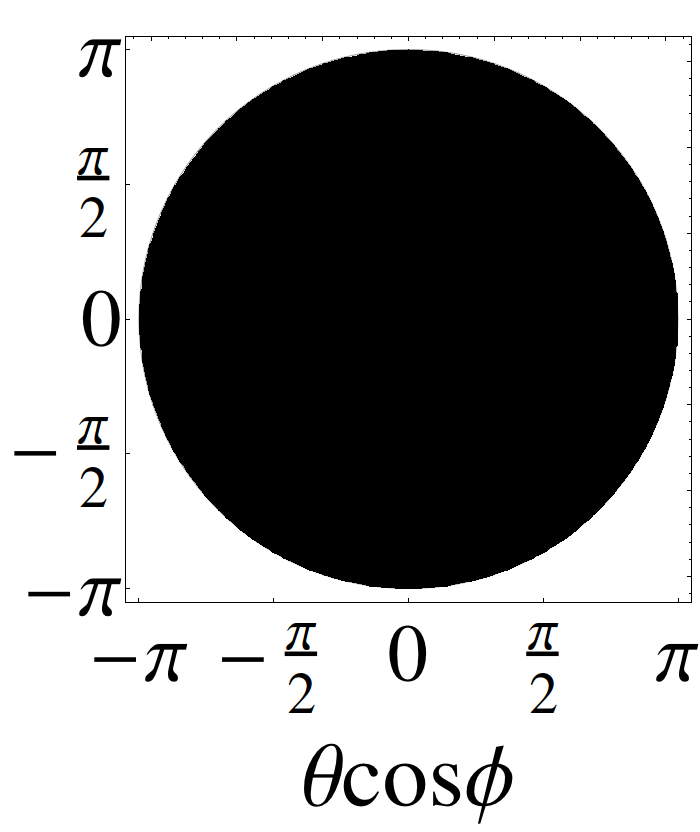}\\
 &\\
 \end{tabular}
\caption{Panels a), b) and c): SDoS for three different couplings ($\gamma/\gamma_+=0.8,1.4$ y $2.6$ respectively)  for  every $j=1,2,...,\mathcal{N}/2$,  in the case $\mathcal{N}=40$ with $\delta=0.2$. The selected couplings correspond to the three different regimes a)$ \gamma<\gamma_+$, b) $\gamma_+<\gamma<\gamma_-$ c) $\gamma_-<\gamma$. The arrow signals the descending order of $j$. The colors in the curves indicate the different regimes in the SDoS discussed in the text. For the larger coupling [panel c)] the four regimes are present in the largest values of $j$. Panel d): For the largest pseudospin ($j=\mathcal{N}/2$) of this latter case, the topology of the pseudospin  available volume is shown for  representative energies ($2E/(\omega_o \mathcal{N})=-2.5,1.4,-0.9$ and $1.2$) of each of this four regimes, using the variable $\theta=\arccos j_z/j$.} 
\label{fig:7}
\end{figure}

For $\gamma<\gamma_{j,+}$, two different energy regimes exist, one for $-j\omega_o\leq E<+j\omega_o$ and a second one for  $+j\omega_o\leq E$, at these large energies  the whole Bloch sphere becomes available.  The SDoS is
\begin{equation}
\frac{\omega}{2j}\nu_{\delta}(E,\mathcal{N},j)=\left\{
\begin{array}{cl}
 \frac{y_{0+}+1}{2}+\frac{1}{\pi}\int_{y_{0+}}^{y_{1+}}F_j(y,\epsilon_j)dy , & \mbox{for}\,\,\, -1\leq\epsilon_j < 1, \\
 1, & \mbox{for}\,\,\,1\leq\epsilon_j.
\end{array}
\right.
\end{equation}
where $\epsilon_{j}=\frac{E}{\omega_{0}j}$ and  we have used the following definitions
\begin{equation}
\fl F_j(y,\epsilon_j)=\arccos \sqrt{\left(\left(\frac{\gamma}{\gamma_{j,+}}\right)^{2}-\left(\frac{\gamma}{\gamma_{j,-}}\right)^{2}\right)^{-1}\left[\frac{2\left(y-\epsilon_{j}\right)}{\left(1-y^{2}\right)}-\left(\frac{\gamma}{\gamma_{j,-}}\right)^{2}\right]},
\end{equation}
\begin{eqnarray}
y_{0\pm}&=&-\left(\frac{\gamma_{j,-}}{\gamma}\right)^{2}\pm\left(\frac{\gamma_{j,-}}{\gamma}\right)\sqrt{2\left(\epsilon_{j}-\epsilon_{j-}\right)},
\mbox{   \ \ \ \   and  \ \ \ \      } \\ \nonumber
y_{1\pm}&=&-\left(\frac{\gamma_{j,+}}{\gamma}\right)^{2}\pm\left(\frac{\gamma_{j,+}}{\gamma}\right)\sqrt{2\left(\epsilon_{j}-\epsilon_{j+}\right)},
\end{eqnarray}
with
$$
\epsilon_{j\pm}=\frac{E_{j,e}^{\pm}}{j\omega_{0}}=-\frac{1}{2}\left[\left(\frac{\gamma_{j,\pm}}{\gamma}\right)^{2}+\left(\frac{\gamma}{\gamma_{j,\pm}}\right)^{2}\right].
$$
 
For $\gamma_{j,+}\leq \gamma<\gamma_{j,-}$, the normal to superradiant QPT has already occurred, and  a new energy regime appears, which is defined by the interval $E_{j,e}^+\leq E < -j\omega_o$. The  other two intervals are the same as in the previous case ($-j\omega_o\leq E<+j\omega_o$ and  $+j\omega_o\leq E$). The SDoS is
\begin{equation}
\frac{\omega}{2j}\nu_{\delta}(E,\mathcal{N},j)=\left\{
\begin{array}{cl}
\frac{1}{\pi}\int_{y_{1-}}^{y_{1+}} F_j(y,\epsilon_j) dy , & \mbox{for}\,\,\,\epsilon_{j+} \leq \epsilon_j < -1\\
\frac{y_{0+}+1}{2}+\frac{1}{\pi}\int_{y_{0+}}^{y_{1+}}F_j(y,\epsilon_j)dy , & \mbox{for}\,\,\, -1\leq\epsilon_j < 1, \\
 1, & \mbox{for}\,\,\,1\leq\epsilon_j.
\end{array}
\right.
\end{equation}

Finally,  for $\gamma_{j,-}\leq \gamma$ a new intertwined energy  interval appears, yielding four intervals {\bf 1)} $E_{j,e}^+\leq E <E_{j,e}^-$, {\bf 2)} $E_{j,e}^-\leq E <-j\omega_o$, {\bf 3)} $-j\omega_o\leq E<+j\omega_o$, and 4) $+j\omega_o\leq E$. The SDoS is 

\begin{equation}
\fl \frac{\omega}{2j}\nu_{\delta}(E,\mathcal{N},j)=\left\{
\begin{array}{cl}
\frac{1}{\pi}\int_{y_{1-}}^{y_{1+}} F_j(y,\epsilon_j) dy , & \mbox{for}\,\,\,\epsilon_{j+} \leq \epsilon_j < \epsilon_{j-}\\
\frac{y_{0+}-y_{0-}}{2}+\frac{1}{\pi}\int_{y_{1-}}^{y_{0-}}F_j(y,\epsilon_j)dy +& \mbox{for}\,\,\,\epsilon_{j-} \leq \epsilon_j < -1\\
+\frac{1}{\pi}\int_{y_{0+}}^{y_{1+}}F_j(y,\epsilon_j)dy, & \\
\frac{y_{0+}+1}{2}+\frac{1}{\pi}\int_{y_{0+}}^{y_{1+}}F_j(y,\epsilon_j)dy , & \mbox{for}\,\,\, -1\leq\epsilon_j < 1, \\
 1, & \mbox{for}\,\,\,1\leq\epsilon.
\end{array}
\right.
\end{equation}

The energies separating two contiguous energy intervals define the so called Excited-State Quantum Phase Transitions (ESQPTs), because at these energies  critical changes in the properties of the SDoS and in the topology of the available phase space (see Fig.\ref{fig:7}) are observed. For $\gamma<\gamma_{j,+}$ only one ESQPT is observed at $E=j\omega_0$. For $\gamma_{j,+}\leq \gamma<\gamma_{j,-}$, two ESQPTs are present, one at $E=-j\omega_0$,  called \emph{dynamical} \cite{Bas14}, and other at the same energy as before $E=j\omega_0$, which was called \emph{static}. For the last case $\gamma_{j,-}\leq \gamma$, in addition to the two  previous ESQPTs, a third ESQPT at critical energy $E=E_{j,e}^-$, occurs. In Fig.\ref{fig:7} the SDoS for every pseudospin, $j=1,2,...,\mathcal{N}/2$, are  plotted for three different couplings ($\gamma<\gamma_+$, $\gamma_+<\gamma<\gamma_-$, and $\gamma_-<\gamma$). The colors of the curves indicate the different energy regimes in the SDoS. The number of colors in the curves gives the number of ESQPTs in the corresponding SDoS: {\bf a)} two colors (blue and black) one ESQPT, {\bf b)} three colors (red, blue and black) two  ESQPTs and {\bf c)}  four colors (red, orange, blue and black) three ESQPTs. Observe that, for the last case ($\gamma_-<\gamma$), the three kind of SDoS are obtained when the pseudo-spin $j$ is varied, 3-ESQPTs SDoS for large $j$,  2-ESQPTs SDoS for intermediate $j$ and 1-ESQPT SDoS for small $j$. 
It is interesting to observe (see  Fig.\ref{fig:6}) that the region in the energy vs coupling space corresponding to $E_{j,e}^-\leq E\leq -j\omega_o$ (the orange part of   the SDoS curves) coincides with the region where the function  $\phi_\delta(u_+,u_-)$ of the canonical ensemble present  two saddle points at $(u_+,u_-)=(0,\pm u_-^{m})$, i.e. the region we have, already identified as the second superradiant phase that could be relevant in the finite size  ($\mathcal{N}<\infty$) case. 

With this, we have obtained the first necessary ingredient  
in order to calculate the number of states $\Omega(E,\mathcal{N})$. Aside from finding the thermal meaning of the QPT, we have deduced the SDoS for each pseudo-spin $\nu(E,\mathcal{N},j)$. While the SDoS results are interesting by themselves, we will show below that their contribution, including their critical properties (the ESQPTs), are completely negligible in the thermodynamic limit. However, they could give rise to physical manifestations when finite number corrections are considered. In the next section we study the properties of the multiplicities $Y(\mathcal{N},j)$, whose properties allow to justify the previous assertions.  


\subsection{Multiplicities}

Now, we explore the behavior of the multiplicities $Y(\mathcal{N},j)$, we will show that its thermodynamical ($\mathcal{N}\rightarrow\infty$) properties completely determine the behavior of the number of states for a given energy, and consequently the entropy of the system. We consider a set of $\mathcal{N}$ distinguishable particles or qubits, then the degeneracy of  each pseudospin $j$, the number of {\it physically distinguishable states},  is given by \cite{Ross84,Wilms12}
\begin{equation}
\fl Y(\mathcal{N},j)={\mathcal{N} \choose \frac{\mathcal{N}}{2}-j}-{\mathcal{N} \choose \frac{\mathcal{N}}{2}-j-1}=\left(2j+1\right)\left[\frac{\mathcal{N}!}{\left(\frac{\mathcal{N}}{2}-j\right)!\left(\frac{\mathcal{N}}{2}+j+1\right)!}\right].
\end{equation}
It is interesting to observe that, if we consider a set of $\mathcal{N}$ bosons, only the completely symmetric representation of the collective pseudospin, which corresponds to the maximum value of $j=\mathcal{N}/2$, has to be considered. Consequently, the number of states is given by $\mathcal{N}+1$, and the entropy goes zero in the thermodynamic limit giving no thermodynamic observables \cite{Apa12}.
 
\begin{figure}
\centering{
\begin{tabular}{c c}
a)&b)\\
\includegraphics[angle=0,width=0.5\textwidth]{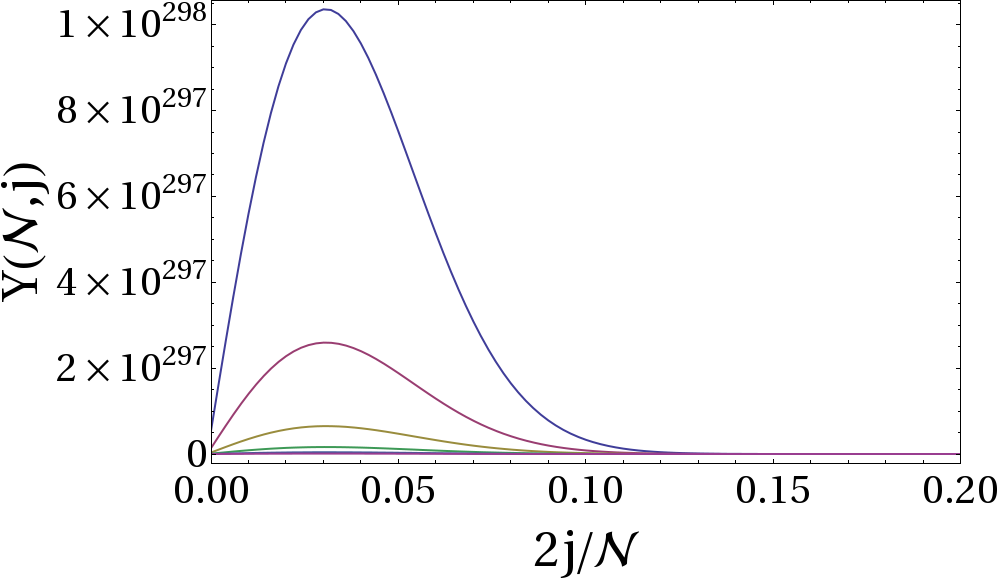} &
\includegraphics[angle=0,width=0.5\textwidth]{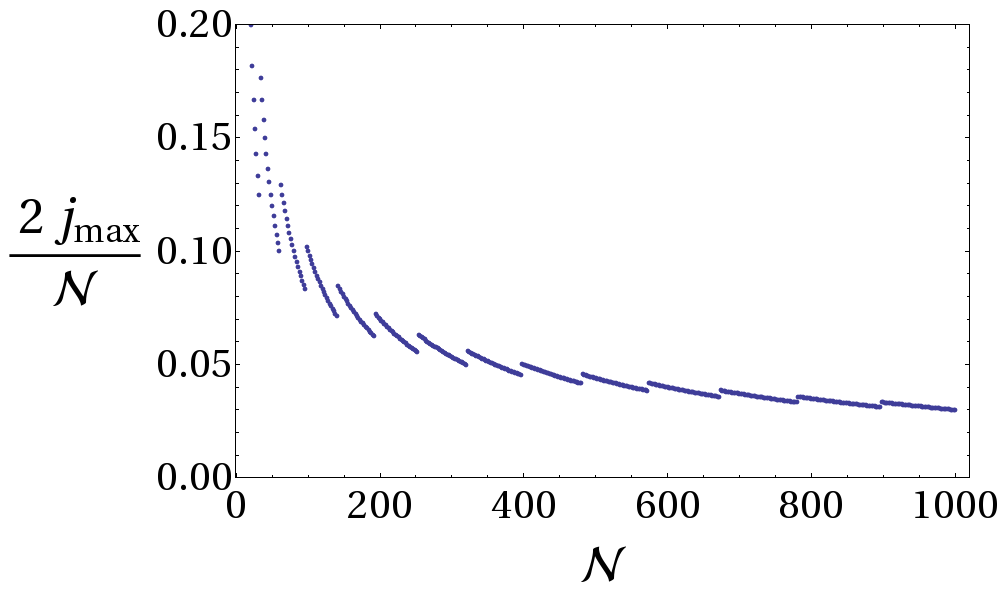} \\
\end{tabular}}
\caption{a): $Y(\mathcal{N},j)$ as a function of $2j/\mathcal{N}$ for $\mathcal{N}=1000,998,...,990$ (from upper to lower curve). b) $j$ that maximize  $Y(\mathcal{N},j)$ divided by $\mathcal{N}/2$ ($z_{max}=2j_{max}/\mathcal{N}$) as a function of $\mathcal{N}$. Observe that the $z=2j/\mathcal{N}$ value that maximize  $Y(\mathcal{N},j)$ decreases as $\mathcal{N}$ grows}
\label{fig:8}
\end{figure}

In Fig. \ref{fig:8} we plot the multiplicities as a function of $z\equiv 2 j/\mathcal{N}\in [0,1]$, for different number of two-level atoms $\mathcal{N}$.  

We can observe that there is a dominant pseudospin whose ratio $2j/\mathcal{N}$  goes to zero as $\mathcal{N}\rightarrow\infty$. The multiplicities represent the statistical weight for a given $j$-sector and, as we are interested on thermal observables, we want to know the behavior of $Y(\mathcal{N},j)$ in the thermodynamical limit. In order to do so we employ the Stirling approximation 
\begin{eqnarray}
 \ln\left(Y\right)\simeq & \ln\left(2j+1\right)+1+\mathcal{N}\,\ln\mathcal{N}+\\ \nonumber
&-\left(\frac{\mathcal{N}}{2}-j\right)\,\ln\left(\frac{\mathcal{N}}{2}-j\right)-\left(\frac{\mathcal{N}}{2}+j+1\right)\,\ln\left(\frac{\mathcal{N}}{2}+j+1\right).
\end{eqnarray}
If we express the previous formula in terms of variable $z\equiv 2j/\mathcal{N}\in[0,1]$, we obtain 
\begin{eqnarray}
\fl \ln\left(Y\right)&\simeq \ln\left(2z+\frac{2}{\mathcal{N}}\right)+1+\\ \nonumber
\fl &+\frac{\mathcal{N}}{2}\left[\ln(4)-(1-z)\,\ln(1-z)-\left(1+z+\frac{2}{\mathcal{N}}\right)\,\ln\left(1+z+\frac{2}{\mathcal{N}}\right)\right].
\end{eqnarray}
Neglecting terms of order less than $\mathcal{N}$, we finally have, 
\begin{equation} \label{eqn:mul}
Y(\mathcal{N},z=2j/\mathcal{N})\approx \left(\frac{4}{(1-z)^{1-z}(1+z)^{1+z}}\right)^{\mathcal{N}/2}.
\end{equation}

\begin{figure}
\centering{
\begin{tabular}{c c}
a)&b)\\
\includegraphics[angle=0,width=0.5\textwidth]{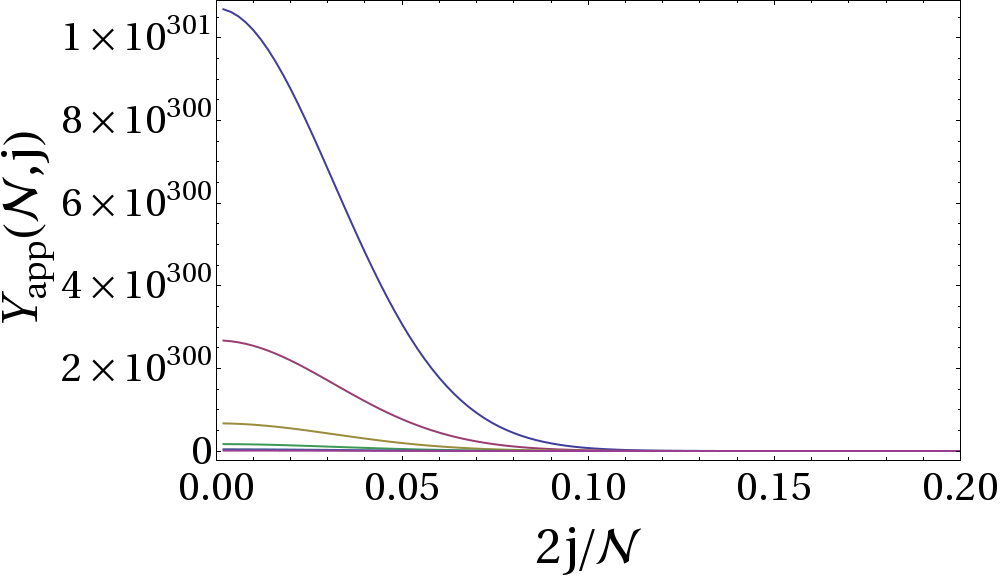} 
&\includegraphics[angle=0,width=0.45\textwidth]{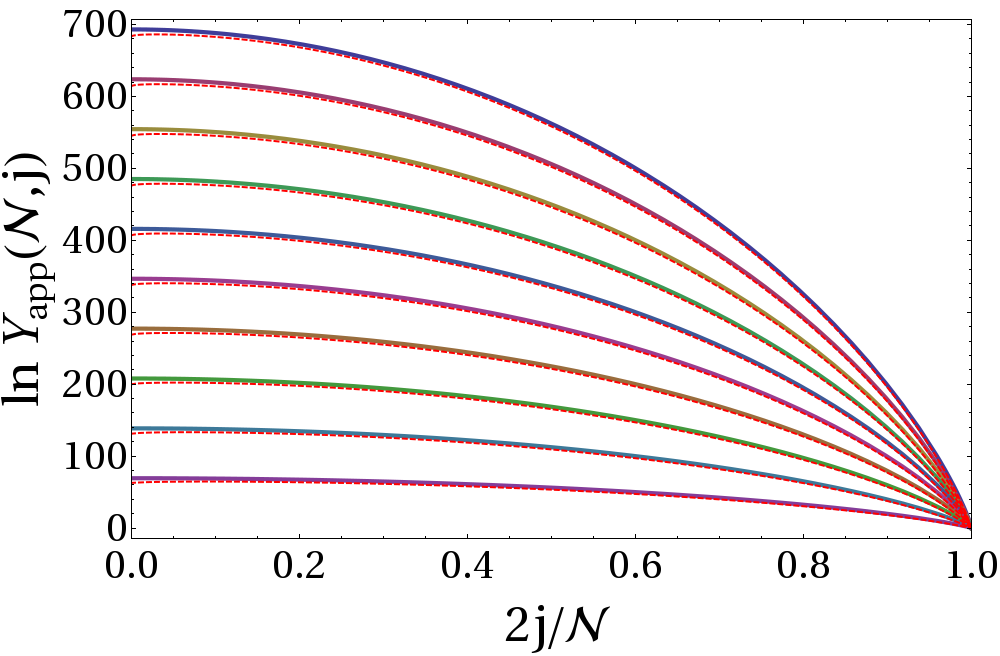} \\
\end{tabular}}
\caption{a): Approximated $Y(j,z)$ as  a function of $z=2j/\mathcal{N}$ for the same cases of Fig.\ref{fig:8} $\mathcal{N}=1000,998,...,990$, from upper to lower curve respectively. The curves show clearly, that, in the approximation,  $Y(j,z)$   is a monotone decreasing function of $z$. b): Solid lines depict the logarithm of approximated  $Y(j,z)$ for $\mathcal{N}=1000,900,800,...,100$  (from upper to lower curve respectively), the dashed red curves depict the exact results.}
\label{fig:9}
\end{figure}

In the thermodynamical limit, the multiplicity is a monotone decreasing function of the variable $z$ (see  Fig.\ref{fig:9}), whose   maximum value,  at $z=0$ (corresponding to lowest  $j=0$ pseudospin), is $Y(\mathcal{N},0)=2^\mathcal{N}$.   Furthermore, in the same  limit, all the multiplicities of the larger pseudo-spins $z+\epsilon=2(j+\Delta j)/\mathcal{N}$ (with $\epsilon,\Delta j>0$) are negligible respect to the multiplicity of the given  pseudospin $z=2j/\mathcal{N}$   
$$
\lim_{\mathcal{N}\rightarrow\infty}\frac{Y(\mathcal{N},z+\epsilon)}{Y(\mathcal{N},z)}=0, \ \ \ \ \mbox{for all } \epsilon>0.
$$ 
Therefore, since the SDoS calculated before grows linearly with $j$, for a fixed energy $E$ the number of states will be entirely dominated by the multiplicity of the lowest pseudo-spin compatible with that energy.  From the study of the lowest energy for every $j$, we know that, for any coupling, the minimal energy of the pseudo-spin $j$ increases as $j$ decreases (as it can be seen in Fig. \ref{fig:5}). Consequently,  for a given coupling and energy ($E<0$) only the largest pseudo-spins (those that satisfy $E_{j}^{gs}\leq E$) are available and the multiplicity of the smallest one, $j_m$,  determines the thermodynamics of the system. Then, in the thermodynamic limit,  the leading contribution to the number of states for a given energy $E$ is given by the  multiplicity $Y(\mathcal{N},z_m=2 j_{m}/\mathcal{N})$ 
\begin{equation}
\ln\left[\Omega(E,\mathcal{N})\right]=\ln\left[\sum_{j=0}^{\mathcal{N}/2}Y(\mathcal{N},j)\,\nu(E,\mathcal{N},j)\Delta E\right]\simeq \ln\left[Y(\mathcal{N},z_m)\right],
\end{equation}
where $Y(\mathcal{N},z_m)$ is the multiplicity (\ref{eqn:mul}) of the smallest  $j$-pseudospin available  for the energy $E$. The smallest pseudo-spin compatible with a given energy $E<0$, can be obtained   by solving, for $j$, the equation
$$
E_{j}^{gs}(\gamma,\delta)=E,
$$
where $E_{j}^{gs}(\gamma,\delta)$  is  the lowest energy for given $j$,  Eq.(\ref{eqn:gse}). The  solution, $j_m(E)$, to the previous equation is given and discussed in the following subsection.

\subsection{Entropy}

With the  results of the previous subsections, we have all the necessary ingredients to calculate the thermodynamics of the generalized Dicke model from a microcanonical ensemble approach. As it was discussed before, the number of states for a given energy $E$, and consequently the entropy is determined, in the thermodynamic limit, by the multiplicity of the minimal pseudo-spin  $z_m=2 j_{m}/\mathcal{N}$, compatible with $E$
$$
S(E)=k_B \ln \left[Y(\mathcal{N},z_m)\right],
$$
where $z_m$ is given by the solution of  $E_{j}^{gs} (\gamma,\delta)=E$. From  Eq. (\ref{eqn:gse}) and using Fig.\ref{fig:5} as a guide, it is clear that the previous equation takes two different forms depending on the values of coupling and energy. For the case $\gamma<\gamma_+$ or $\gamma\geq \gamma_+$ with $E\geq E_j^{QPT}$ [the already identified  normal phase obtained in the canonical ensemble approach, with  $E_j^{QPT}$  the critical energy given by Eq.(\ref{eqpt})], the equation $E_{j}^{gs}(\gamma,\delta)=E$ takes the simple form $E=-j\omega_0$, whose solution is clearly $j_m=-E/\omega_0$, which implies
 \begin{eqnarray}\label{zm1}
\fl z_m=2 j_m/\mathcal{N}&=&-2 E/(\omega_o\mathcal{N})\\ \nonumber 
\fl &&\mbox{for the normal phase ($\gamma<\gamma_+$ or $\gamma\geq \gamma_+$ with $E\geq E_j^{QPT}$)}.
\end{eqnarray}
For the case $\gamma\geq \gamma_+$ with $E < E_j^{QPT}$ (the already identified superradiant phase) the equation  $E_{j}^{gs}(\gamma,\delta)=E$ becomes 
$$
E=-\frac{j\omega_o}{2}\left[\left(\frac{\gamma_{j,+}}{\gamma}\right)^2+\left(\frac{\gamma}{\gamma_{j,+}}\right)^2\right]=-\frac{j\omega_o}{2}\left[\left(\frac{\mathcal{N}}{2j}\right)\left(\frac{\gamma_{+}}{\gamma}\right)^2+\left(\frac{2j}{\mathcal{N}}\right)\left(\frac{\gamma}{\gamma_{+}}\right)^2\right],
$$
whose solution is given by
\begin{eqnarray}\label{zm2}
\fl z_m&=&\frac{2 j_m}{\mathcal{N}}=\left(\frac{\gamma_{+}}{\gamma}\right)\sqrt{-\left(\frac{4E}{\omega_o \mathcal{N}}+\left(\frac{\gamma_{+}}{\gamma}\right)^{2}\right)}\\ \nonumber
\fl &&\mbox{for the superradiant  phase ($\gamma\geq \gamma_+$ with $E < E_j^{QPT}$) }.
\end{eqnarray}
The minimal pseudospin $z_m$ is equal to the function $\mathcal{E}_{\delta}$ (\ref{funu}) defined in the canonical ensemble approach to express the entropy as a function of the energy. Gathering all the previous results, is  very easy to prove that the entropy from the microcanonical approach is exactly the same obtained previously from the canonical ensemble. The entropy per particle is given by 
\begin{eqnarray}
\fl \frac{\mathcal{S}(\epsilon)}{k_{B}}=\frac{S(E,\mathcal{N})}{k_{B}\mathcal{N}}&=&\frac{1}{\mathcal{N}}\ln\left[Y(\mathcal{N},z_{m})\right]=
\frac{1}{\mathcal{N}}\ln\left[\left(\frac{4}{(1-z_m)^{1-z_m}(1+z_m)^{1+z_m}}\right)^{\mathcal{N}/2}\right]\nonumber\\
&=&\ln(2) -\frac{1}{2}\left(1+z_{m}\right)\,\ln\left(1+z_{m}\right)-\frac{1}{2}\left(1-z_{m}\right)\,\ln\left(1-z_{m}\right),
\end{eqnarray}
which is equal to the canonical result of Eq.(\ref{entrocan}) remembering that $z_{m}=\mathcal{E}_{\delta}$ and that the internal energy $\mathcal{U}_\delta=E$. 
Therefore, we have solved the thermodynamics of the generalized Dicke model in the micro-canonical ensemble. 

Let us discuss briefly the case $E\geq 0$. As the lowest energies for every $j$ satisfy  $E_{j}^{gs}\leq 0$, all the pseudo-spins are available for these energy values. Therefore, for $E\geq 0$ the number of states is approximated by the multiplicity of the smallest pseudospin $j=0$, which, as it was noted before, is approximated by $\Omega(E\geq0,\mathcal{N}\simeq Y(\mathcal{N},0)=2^{\mathcal{N}}$.  This  multiplicity is equal to  the dimension of the atomic subspace and independent on energy, consequently the entropy becomes constant for $E>0$ and, since  $\partial S/\partial E=1/T=0$, this energy region is unreachable at finite temperature. The same result obtained previously from   the canonical ensemble approach to the  thermodynamics.  Previously,   in \cite{Jaw85}, a gaussian approximation to the micro-canonical ensemble was used to study the Dicke Hamiltonian.  There it was concluded that the internal energy $\mathcal{U}=0$  is  another thermal phase transition. From our results  is clear,  however, that $\mathcal{U}=0$ is simply the infinite temperature limit. 

\subsection{Critical temperature and internal energy}

Other thermodynamical observables can be obtained form the microcanonical approach.
 We calculate the temperature from the entropy, 
\begin{eqnarray}
\beta&=\frac{1}{k_{B}T}=\frac{1}{k_{B}}\frac{dS}{dE}=\frac{2}{\omega_{0}k_{B}}\frac{d\mathcal{S}}{d\epsilon}=\\ \nonumber
&=-\frac{1}{\omega_{0}}\,\ln\left(\frac{1+z_{m}}{1-z_{m}}\right)\frac{dz_{m}}{d\epsilon}=-\frac{2}{\omega_{0}}\,\mbox{arctanh}\left(z_{m}\right)\frac{dz_{m}}{d\epsilon},
\end{eqnarray}
where the derivative of $z_m$ is
\begin{equation}
\fl \frac{dz_{m}}{d\epsilon}=\left\{\begin{array}{cc}
-\left(\frac{\gamma_{+}}{\gamma}\right)\left[-\left(2\epsilon+\left(\frac{\gamma_{+}}{\gamma}\right)^{2}\right)\right]^{-1/2} & \mbox{for the superradiant phase} \\ \nonumber
& \mbox{ ($\gamma\geq \gamma_+$ with $E < E_j^{QPT}$) }\\
-1 & \mbox{for the normal phase} \\ \nonumber
& \mbox{($\gamma<\gamma_+$ or $\gamma\geq \gamma_+$ with $E\geq E_j^{QPT}$)} \\
0 & \mbox{for}\,\,\, E\geq 0,
\end{array}\right.
\end{equation}
with $\epsilon=2E/(\omega_o\mathcal{N})$.
Then, the temperature as a function of the energy is, 
\begin{equation} \label{eqn:temmic}
\fl \beta=\left\{\begin{array}{cc}
\frac{2}{\omega_{0}}\left(\frac{\gamma_{+}}{\gamma}\right)\left[-\left(2\epsilon+\left(\frac{\gamma_{+}}{\gamma}\right)^{2}\right)\right]^{-1/2}\times & \mbox{for the superradiant phase}\\
\times \mbox{arctanh}\left\{\left(\frac{\gamma_{+}}{\gamma}\right)\left[-\left(2\epsilon+\left(\frac{\gamma_{+}}{\gamma}\right)^{2}\right)\right]^{1/2} \right\} & \\
\frac{2}{\omega_{0}}\,\mbox{arctanh}(-\epsilon) & 
\mbox{for the normal phase} \\
\end{array}\right.
\end{equation}

By evaluating the temperature  at  the critical energy, $\epsilon_{c}$,  we obtain the critical temperature, which separates the normal and superradiant thermal phases
\begin{equation}
\beta_{c}^{+}=\frac{2}{\omega_{0}}\mbox{arctanh}\left(\left(\frac{\gamma_{+}}{\gamma}.\right)^{2}\right).
\end{equation}
We have recovered the critical temperature of the finite-temperature super radiant phase transition obtained previously from the canonical ensemble. Besides, as it was already mentioned,  we observe that for $\epsilon\geq 0$ the derivative $\frac{dz_{m}}{d\epsilon}=0$, consequently, $\beta\rightarrow 0$ and this energy range is unavailable for finite temperatures. 

Finally, from (\ref{eqn:temmic}) we can recover the internal energy in the micro canonical ensemble. For example, in the normal phase the energy is
\begin{equation}
\epsilon=-\tanh\left(\eta\right),
\end{equation}
which agrees with our calculations in the canonical ensemble.

Before concluding, let us discuss the critical properties of the generalized Dicke energy spectrum and their relation with the thermodynamical critical phenomena, under the light of our microcanonical results.  As it was shown, the thermodynamical properties of the model are entirely given by the properties of the lowest energy state at each pseudospin $j$. Particularly, the thermal  phase transition line between the normal and superradiant phase, is  the aggregated of the QPTs of each pseudospin $j=0,..,\mathcal{N}/2$. On the other hand, the other critical phenomena observed in the energy spectrum, the so called ESQPTs, since they occur at energies larger than the minimal energy of each pseudo-spin $j$ have no effect or manifestation in the thermal properties. One of them, the called {\it  static} ESQPT  occuring at energy $E=\omega_0 j$,  belongs to a forbidden thermodynamic energy region ( $E\geq 0$). The other two ESQPTs, the one occurring at $E=-\omega_o j$ and that at $E=E_{j,e}^-$,  even if they are in an energy range thermally available, their contributions to the thermodynamics are negligible, and their effects disappear completely in the thermodynamical limit.  Nonetheless, it would be interesting  to determine whether the ESQPTs have any effect or  manifestation  in the finite size corrections to the thermodynamical ($\mathcal{N}\rightarrow\infty$) limit, which goes beyond scope of the present contribution. Other interesting topic in the study of the finite size corrections, would be the comparison between a canonical and a microcanonical approach. In the thermodynamical limit, this work has explicitly shown that both descriptions give exactly the same results.            

\section{Conclusions}

We have solved the thermodynamics of the generalized Dicke model both in the canonical and, for the first time, in the microcanonical ensemble. In order to calculate the microcanonical ensemble we employed a semi-classical approximation for the Density of States.

We showed results for all the relevant observables, which let us in a simple way recover the results for interesting temperature limits, $T\rightarrow 0$, $T\rightarrow \infty$ and $T\rightarrow T_c$. We 
have demonstrated that the results for both ensembles agree in the thermodynamical limit, and, in this way we linked the point of view of the canonical statistical ensambles and the perspective of isolated quantum systems. Besides, we obtained expression for  the semiclassical DoS for the extended Dicke model. All of these calculations could help to study problems with a tunable parameter between an integrable system (Tavis-Cummings) and a non-integrable one (Dicke).

Like the Dicke and Tavis-Cummings models, the generalized Dicke model studied here exhibits only two thermal phases in the thermodynamical limit, a normal and a superradiant phase. However, unlike the Dicke or Tavis-Cummings models, we identify a region which potentially could give rise to a second superradiant phase. However, this second superradiant phase cannot be an equilibrium state. From the thermodynamic point of view, it corresponds to a saddle point, i. e. a non stable phase. From the semi-classical point of view, it is not the minimum of the classical energy surface. The minimum corresponds to the first superradiant phase. The second superradiant phase could have observable effects only in the finite size case. 

We connected in a simple way the thermal phase transition in the generalized Dicke model with the QPTs of the lowest energy states of each subspace of pseudospin $j$, by calculating, for each subspace,  their degeneracies and  semi-classical lowest
energies and densities of states. We found that the curve in the energy-coupling space where the QPT for each subspace $j$ takes place reproduces  the curve of the critical energy corresponding to the critical temperature in the phase diagram, i.e., to the thermal phase transition. Then, the superradiant QPT is the thermal superradiant phase transition. So, we have related these critical phenomena in the spectrum with its thermal counterpart. 

The excited states of every $j$ are thermodynamically inaccessible, and the thermodynamical properties of the model are entirely given by the lowest energy states of each pseudospin $j$. Consequently  the critical phenomena observed in the excited energy spectrum, the so called ESQPTs, have no effect and manifestation in the thermal properties. However they could be of interest in a finite size study of the model \cite{PerezArmando}. The region of positive energies was also shown to be thermally inaccesible.

The formalism presented here is immediately applicable to other systems formed by a set of $\mathcal{N}$ identical  few level atoms   whose   Hamiltonians are  expressed in terms of collective operators satisfying a given algebra.  Example of this is the Lipkin-Meshkov-Glick model whose Hamiltonian is expressed in terms of $SU(2)$ operators. We hope the formalism presented here  could help to  understand the relationship between  the  critical properties of the quantum spectrum   with the  thermal critical phenomena. Also, this approach could help the developing of techniques to study finite temperature problems in isolated quantum many-body systems. 

\section{Acknowledgements}
We thank A. Rela\~no, P. P\'erez-Fern\'andez, P. Stranksy and M. Kloc for fruitful discussions which lead to the development of this work and its enrichment. Also, we show our gratitude to P. Cejnar for the support and insight, as well as, for the reading of the manuscript. M. Bastarrachea-Magnani wants to thank R. Rossignoli for his interest on the subject and his pertinent advices.
This work has received partial economical support from SEP-Conacyt, and from RedTC-Conacyt, Mexico. 

\appendix


\section{Calculation of the partition function}
In this appendix we calculate the canonical partition function, Eq.(\ref{zfunc}), 
\begin{equation}
\fl \mathcal{Z}_{\delta}(T,\mathcal{N})=\int \frac{d^{2}\alpha}{\pi}\sum_{s_{1}=\pm}\sum_{s_{2}=\pm}\cdot\cdot\cdot\sum_{s_{\mathcal{N}}=\pm} \langle\alpha|\langle s_{1}|\langle s_{2}|\cdot\cdot\cdot\langle s_{\mathcal{N}}| e^{-\beta H_{D,\delta}} |\alpha\rangle |s_{1}\rangle|s_{2}\rangle\cdot\cdot\cdot|s_{\mathcal{N}}\rangle.
\end{equation}

Following \cite{WH73,CGW73} the field trace is, 
\begin{eqnarray}
\fl &&\langle \alpha| e^{-\beta H_{D,\delta}} |\alpha\rangle=\exp\left\{\sum_{k=1}^{\mathcal{N}}H_{D,\delta}^{k}(\alpha)\right\}=\\ \nonumber
\fl &&=\exp\left\{\sum_{k=1}^{\mathcal{N}}\left[\omega\frac{|\alpha|^{2}}{\mathcal{N}}+\frac{\omega_{0}}{2}\sigma_{z}^{k}+\frac{\gamma}{2\sqrt{\mathcal{N}}}\left((1+\delta)Re(\alpha)\sigma_{x}^{k} - (1-\delta)Im(\alpha)\sigma_{y}^{k}\right)\right]\right\}.
\end{eqnarray}
Then, we arrange the sum taking advantage of the independence between the atoms,  
\begin{eqnarray}
\mathcal{Z}_{\delta}(T,\mathcal{N})&=&\int \frac{d^{2}\alpha}{\pi} e^{-\mathcal{N}\beta \omega \frac{|\alpha|^{2}}{\mathcal{N}}} \prod_{k=1}^{\mathcal{N}}\sum_{s_{k}=\pm}\langle s_{k}| e^{-\beta h_{\delta,k}(\alpha)}|s_{k}\rangle=\\ \nonumber
&=& \int \frac{d^{2}\alpha}{\pi} e^{-\mathcal{N}\beta \omega \frac{|\alpha|^{2}}{\mathcal{N}}} \left[\sum_{s=\pm}\langle s| e^{-\beta h_{\delta}(\alpha)}|s\rangle\right]^{\mathcal{N}}.
\end{eqnarray} 
where, 
\begin{equation}
h_{\delta}(\alpha)=\frac{\omega_{0}}{2} \sigma_{z}+\frac{\gamma}{2\sqrt{\mathcal{N}}}\left[(1+\delta)Re(\alpha)\sigma_{x} - (1-\delta)Im(\alpha)\sigma_{y}\right].
\end{equation}
In order to calculate the atomic trace we find the eigenstates and egenvalues of  $h_{\delta}$. In the $\sigma_{z}$ basis we have,
\small
\begin{eqnarray}
\fl & h_{\delta}(\alpha)=\left( \begin{array}{c c} 
\frac{\omega_{0}}{2} & \frac{\gamma}{\sqrt{\mathcal{N}}}\left[(1+\delta) Re(\alpha)+(1-\delta)i Im(\alpha)\right] \\
\frac{\gamma}{\sqrt{\mathcal{N}}}\left[(1+\delta) Re(\alpha)-i(1-\delta)i Im(\alpha)\right]  & -\frac{\omega_{0}}{2} \\
\end{array}
\right)
\end{eqnarray} 
\normalsize
The eigenvalue equation is, 
\begin{equation}
-(\frac{\omega_{0}}{2} - \lambda)(\frac{\omega_{0}}{2} + \lambda) - \frac{\gamma^{2}}{\mathcal{N}}\left[(1+\delta)^{2} Re^{2}(\alpha)+(1-\delta)^{2} Im^{2}(\alpha)\right]=0.
\end{equation}
The eigenvalues $\lambda_{\delta,\pm}(\alpha)$ are, 
\begin{equation}
\lambda_{\delta,\pm}(\alpha)=\pm\sqrt{\frac{\omega_{0}^{2}}{4}+\frac{\gamma^{2}}{\mathcal{N}}\left[(1+\delta)^{2} Re^{2}(\alpha)+(1-\delta)^{2} Im^{2}(\alpha)\right]},
\end{equation}
and the eigenstates,
\begin{equation} \label{ch3eigen}
|\lambda_{\delta,\pm}\rangle=\sqrt{\frac{\lambda_{\delta,\pm}-a}{2\,\lambda_{\delta,\pm}}}\left(\begin{array}{c} \frac{\frac{\omega_{0}}{2}+\lambda_{\delta,\pm}(\alpha)}{\frac{\gamma}{\sqrt{\mathcal{N}}}\left[(1+\delta)Re(\alpha)-(1-\delta)i\,Im(\alpha) \right]} \\ 1  \\ \end{array}\right).
\end{equation}
Then, evaluating the trace for the atomic sector in this new basis we have, 
\begin{eqnarray} \label{zint}
\mathcal{Z}_{\delta}(T,\mathcal{N})&=&\int \frac{d^{2}\alpha}{\pi} e^{-\beta \omega |\alpha|^{2}} \left(\sum_{\lambda=\lambda_{\delta,\pm}(\alpha)} e^{-\beta \lambda}\right)^{\mathcal{N}}=\\ \nonumber
&=&\int \frac{d^{2}\alpha}{\pi} e^{-\beta \omega |\alpha|^{2}} \left(e^{-\beta \lambda_{\delta,+}(\alpha)}+e^{-\beta \lambda_{\delta,-}(\alpha)}\right)^{\mathcal{N}}.
\end{eqnarray} 
We introduce the function $\chi_{\delta}(\alpha)$, defined through 
\begin{equation}
\frac{\omega_{0}}{2}\chi_{\delta}(\alpha)=\lambda_{\delta,+}(\alpha)=-\lambda_{\delta,-}(\alpha),
\end{equation}
in order to express (\ref{zint}) as 
\begin{equation}
\mathcal{Z}_{\delta}(T,\mathcal{N})=\int \frac{d^{2}\alpha}{\pi} \,e^{\,\mathcal{N}\,\phi_{\delta}(\alpha)}.
\end{equation}
Where  
\begin{equation}
\phi_{\delta}(\alpha)= \frac{-\beta\omega |\alpha|^{2}}{\mathcal{N}} + \ln\left\{2\, \cosh\, \left[\frac{\beta\omega_{0}}{2}\, \chi_{\delta}(\alpha)\right]\right\}
\end{equation}
with 
\begin{equation}
\chi_{\delta}(\alpha)=\sqrt{1+\frac{4\gamma^{2}}{\omega_{0}^{2}\mathcal{N}}\left[(1+\delta)^{2} Re^{2}(\alpha)+(1-\delta)^{2} Im^{2}(\alpha)\right]}.
\label{chi}
\end{equation}
In order to calculate the partition function, we write $\frac{\alpha}{\sqrt{\mathcal{N}}}=u_{+}+iu_{-}$ and the integral in terms of these scaled variables $u_{+}$ and $u_{-}$
\begin{equation}
\mathcal{Z}_{\delta}(T,\mathcal{N})=\frac{\mathcal{N}}{\pi}\int_{-\infty}^{\infty}\int_{-\infty}^{\infty}du_{+}\,du_{-}\, \,e^{\,\mathcal{N} \phi_{\delta}(u_{+},u_{-})},
\end{equation}
where the functions $\phi_{\delta}(\alpha)$ and $\chi_{\delta}(\alpha)$ become
\begin{equation}
\phi_{\delta}(u_{+},u_{-})= -\beta\omega(u_{+}^{2}+u_{-}^{2}) + \ln\left\{2\, \cosh\, \left[\frac{\beta\omega_{0}}{2}\, \chi_{\delta}(u_{+},u_{-})\right]\right\},
\end{equation}
and 
\begin{equation}
\chi_{\delta}(u_{+},u_{-})=\sqrt{1+\frac{4\gamma^{2}}{\omega_{0}^{2}}\left[(1+\delta)^{2} u_{+}^{2}+(1-\delta)^{2} u_{-}^{2}\right]}.
\end{equation}


\section{Calculation of observable averages}

As the Pauli operator $\sigma_{\mu}^{l}$ only acts over one of the spins we can separate that expectation value from the rest. 
\begin{equation} \label{ch3eqJs}
\fl \left\langle\frac{J_{\mu}}{\mathcal{N}}\right\rangle=\frac{1}{\mathcal{N}\mathcal{Z}_{\delta}}\sum_{\ell=1}^{\mathcal{N}}\int \frac{d^{2}\alpha}{2\pi}\,e^{-\beta\omega|\alpha|^{2}}\ \sum_{s_{\ell}=\pm} \langle s_{\ell}| \sigma_{\mu}^{\ell} e^{-\beta\,h_{\ell}(\alpha)} |s_{\ell}\rangle
\prod_{k\neq \ell}^{\mathcal{N}}\sum_{s_{k}=\pm}\langle s_{k}|e^{-\beta h_{k}(\alpha)}|s_{k}\rangle.
\end{equation}

For all the spins $k\neq l$ the expectation values are solved in the same way than in Appendiz A. Now we need to pay attention to  $\sigma_{\mu}^{\ell}\,e^{-\beta\,h_{\ell}(\alpha)}$. In order to evaluate it we need to look at the form of the Pauli matrices in the basis $u_{\delta,\pm}$ i. e. we are interested on evaluating the following expression, 
\begin{equation}
\sum_{u_{\ell}=u_{\delta,\pm}}\langle u_{\ell}| \sigma_{\mu}^{\ell}e^{-\beta\,h_{\ell}(\alpha)}|u_{\ell}\rangle.
\end{equation}

Writing the Pauli matrices using the Kronecker delta as, 
\begin{equation}
\sigma_{\mu}=\left(\begin{array} {c c} \delta_{\mu,z} & \delta_{\mu,x}-i\delta_{\mu,y} \\ \delta_{\mu,x}+i\delta_{\mu,y} & -\delta_{\mu,z} \\ \end{array}\right),
\end{equation}
and employing the eigenvectors $|\lambda_{\pm}\rangle$ in Eqn. \ref{ch3eigen} expressed in the $z$ basis, after multiplying the matrices we have, 
\begin{equation}
\fl \langle \lambda_{\delta,\pm}|\sigma_{\mu}|\lambda_{\delta,\pm}\rangle=\frac{1}{2\,\lambda_{\delta,\pm}}\left\{\omega_{0}\delta_{\mu,z}+\frac{\gamma}{\sqrt{\mathcal{N}}}\left[(1+\delta)Re(\alpha)\delta_{\mu,x}-(1-\delta)Im(\alpha)\delta_{\mu,y}\right]\right\}
\end{equation}

Now, we evaluate the matrix elements realizing we have the same element for all $\ell$,
\begin{eqnarray}
\fl && \sum_{\ell=1}^{\mathcal{N}}\left[\langle \lambda_{\ell,\delta,+}|\sigma_{\mu}\,e^{-\beta h_{\delta,\ell}(\alpha)}|\lambda_{\ell,\delta,+}\rangle+\langle \lambda_{\ell,\delta,-}|\sigma_{\mu}\,e^{-\beta h_{\delta,\ell}(\alpha)}|\lambda_{\ell,\delta,-}\rangle\right]=\\ \nonumber
\fl &&=-\frac{2\mathcal{N}\,\sinh\left(\frac{\beta\omega_{0}}{2}\chi_{\delta}(\alpha)\right)}{\omega_{0}\,\chi_{\delta}(\alpha)}\left\{\omega_{0}\delta_{\mu,z}+\frac{\gamma}{\sqrt{\mathcal{N}}}\left[(1+\delta)Re(\alpha)\delta_{\mu,x}-(1-\delta)Im(\alpha)\delta_{\mu,y}.\right]\right\}
\end{eqnarray}

Therefore, the expression of the thermal averages for $J_{\mu}$ is, using Eqn. \ref{ch3eqJs},
\begin{eqnarray}
\nonumber
\left\langle\frac{J_{\mu}}{\mathcal{N}}\right\rangle_{\delta}&=-\frac{1}{\mathcal{Z}_{\delta}}\int \frac{d^{2}\alpha}{2\pi}\,e^{-\beta\omega|\alpha|^{2}}\frac{\tanh\left(\frac{\beta\omega_{0}}{2}\chi_{\delta}(\alpha)\right)}{\omega_{0}\,\chi_{\delta}(\alpha)}\left(2\,\cosh(\frac{\beta\omega_{0}}{2}\chi_{\delta}(\alpha))\right)^{\mathcal{N}}\times\\ 
& \times\left\{\omega_{0}\delta_{\mu,z}+\frac{\gamma}{\sqrt{\mathcal{N}}}\left[(1+\delta)Re(\alpha)\delta_{\mu,x}-(1-\delta)Im(\alpha)\delta_{\mu,y}\right]\right\}.
\end{eqnarray}

Which is finally,
\begin{eqnarray}
\left\langle\frac{J_{\mu}}{\mathcal{N}}\right\rangle_{\delta}&=-\frac{1}{\mathcal{Z}_{\delta}}\int\frac{d\alpha^{2}}{2\pi}\,\frac{\,\tanh\left(\frac{\beta\omega_{0}}{2}\chi_{\delta}(\alpha)\right)}{\omega_{0}\,\chi_{\delta}(\alpha)}\,e^{\mathcal{N}\phi_{\delta}(\alpha)}\times\\ \nonumber
&\times\left\{\omega_{0}\delta_{\mu,z}+\frac{\gamma}{\sqrt{\mathcal{N}}}\left[(1+\delta)\,Re(\alpha)\,\delta_{\mu,x}-(1-\delta)\,Im(\alpha)\,\delta_{\mu,y}\right]\right\}.
\end{eqnarray}
And in the $u_{\pm}$ [$\alpha=\sqrt{\mathcal{N}}(u_+ + iu_-)$] variables 
\begin{eqnarray} \label{eq:jzxy}
\nonumber
\left\langle\frac{J_{\mu}}{\mathcal{N}}\right\rangle_{\delta}&=-\frac{\mathcal{N}}{\mathcal{Z}_{\delta}}\int_{-\infty}^{\infty}\int_{-\infty}^{\infty} \frac{du_{+}\,du_{-}}{2\pi}\,\frac{\,\tanh\left(\frac{\beta\omega_{0}}{2}\chi_{\delta}(u_{+},u_{-})\right)}{\omega_{0}\,\chi_{\delta}(u_{+},u_{-})}\,e^{\mathcal{N}\phi_{\delta}(u_{+},u_{-})}\times\\
&\times\left\{\omega_{0}\delta_{\mu,z}+\gamma\left[(1+\delta)\,u_{+}\,\delta_{\mu,x}-(1-\delta)\,u_{-}\,\delta_{\mu,y}\right]\right\}.
\end{eqnarray}


\section{The second superradiant phase}

As explained in the main text, there are two superradiant phases for certain Hamiltonian parameters and energies, but they cannot coexist thermodynamically and the second superradiant phase is discarded because it corresponds to a non-stable thermal state. For completeness, we present in this appendix the expressions in the canonical ensemble related to this phase, which are analogous to the first superradiant phase. The function $\phi_{\delta}$ evaluated $(u_{-}^{m},0)$ is, 
\begin{equation}
\phi_{\delta}(u_{-}^{m},0)= -\frac{\beta\omega_{0}}{4}\left(\frac{\gamma_{-}}{\gamma}\right)^{2}\left(\chi_{\delta}^{2}-1\right)+\ln\left[2\, \cosh\, \left(\eta\chi_{\delta}\right)\right], 
\end{equation}
The partition function becomes, 
\begin{equation}
\fl \mathcal{Z}_{\delta}(T,\mathcal{N})=\mathcal{O}_{\delta,-}^{s.p.}(\mathcal{N})\,exp\left[\mathcal{N}\left\{-\frac{\eta}{2}\left(\frac{\gamma_{-}}{\gamma}\right)^{2}\left(\chi_{\delta}^{2}-1\right)+\ln\left[2\, \cosh\, \left(\eta\chi_{\delta}\right)\right]\right\}\right].
\end{equation}
Where
\begin{equation}
\fl \mathcal{O}_{\delta,-}^{s.p.}(\mathcal{N})=\frac{1}{\beta\omega}\left|1 - \left(\frac{\gamma_{-}}{\gamma_{+}}\right)^{2}\right|^{-1/2}\left(\frac{\chi_{\delta}^{2}-1}{\chi^{2}_{\delta}}\right)^{-1/2}\left\{1-\eta\,\left[\left(\frac{\gamma}{\gamma_{-}}\right)^{2}-\left(\frac{\gamma_{-}}{\gamma}\right)^{2}\chi_{\delta}^{2}\right]\right\}^{-1/2}_{\mbox{.}}
\end{equation}
The free energy would be
\begin{equation}
-\beta\mathcal{F}(T)=-\frac{\beta\omega_{0}}{4}\left(\frac{\gamma_{-}}{\gamma}\right)^{2}\left(\chi_{\delta}^{2}-1\right)+\ln\left[2\, \cosh\, \left(\eta\chi_{\delta}\right)\right].
\end{equation}
From where we can obtain the energy and heat capacity
\begin{eqnarray}
\fl \mathcal{U}_{\delta}(T)&=-\frac{\omega_{0}}{4}\,\left(\frac{\gamma_{-}}{\gamma}\right)^{2}\left(\chi_{\delta}^{2}+1\right)=-\frac{\omega_{0}}{4}\,\left[\left(\frac{\gamma}{\gamma_{-}}\right)^{2}\tanh^{2}\left(\eta\,\chi_{\delta}\right)+\left(\frac{\gamma_{-}}{\gamma}\right)^{2}\right], \\
\fl C_{\delta}(T)&=k_{B}\eta^{2}\,\mbox{sech}^{2}\left(\eta\chi_\delta\right) \left(\frac{\gamma}{\gamma_{-}}\right)^{4}\frac{\tanh^{2}\left(\eta\,\chi_{\delta}\right)}{1-\left(\frac{\gamma}{\gamma_{-}}\right)^{2}\eta\,\mbox{sech}^{2}\left(\eta\,\chi_{\delta}\right)}.
\end{eqnarray}


\section{Semiclassical approximation for the DoS using coherent states }

In this appendix we derive the leading order expression of the quantum density of states in the semi-classical approximation, using coherent states. We begin with the quantum density of states 
\begin{equation}
\label{a} \nu_Q \left(E\right) = \sum_{n} \delta \left(E - E_{n}\right),
\end{equation}
with $E_n$ the energy spectrum.
We rewrite the Dirac delta function in eq. \ref{a} in its integral form
\begin{equation}
\label{b}\nu_Q \left(E\right) = \frac{1}{2\pi} \sum_{n} \int_{-\infty}^{\infty} e^{i\left(E - E_{n}\right)\tau} d\tau,
\end{equation}

then, in this way it is possible to write the exponential eq. \ref{b} in terms of the eigenstates of the Hamiltonian
\begin{eqnarray} \label{c}
\nu_Q \left(E\right) &=& \frac{1}{2\pi} \sum_{n} \int_{-\infty}^{\infty} \langle{E_{n}}| e^{i\left(E - E_{n}\right)\tau} |{E_{n}}\rangle d\tau \\
&=& \frac{1}{2\pi} \int_{-\infty}^{\infty} \left[ \sum_{n} \langle{E{n}}| e^{i\left(E - E_{n}\right)\tau} |{E_{n}}\rangle \right] d\tau, \nonumber
\end{eqnarray}
and it can be observed we have the trace of the evolution operator multiplied by a phase
\begin{equation}
\label{e}\nu_Q \left(E\right) = \frac{1}{2\pi} \int_{-\infty}^{\infty} e^{i E \tau} {\mbox Tr}\left[e^{-i \widehat{H} \tau}\right] d\tau.
\end{equation}

Given that the trace of the evolution operator eq. \ref{e} is independent of the basis, we can rewrite it in terms of the coherent state basis
\begin{equation}
\label{f}\nu_Q \left(E\right) = \frac{1}{2\pi} \int_{-\infty}^{\infty} e^{iE\tau}\left[ \int \langle{\alpha,z}| e^{-i \widehat{H} \tau} |{\alpha,z}\rangle d^{2}\alpha~d^{2}z\right] d\tau,
\end{equation} 
where $|{\alpha}\rangle$ represents a coherent state associated to the field,  $|{z}\rangle$ represents a coherent state associated to the pseudo-spin, $|{\alpha,z}\rangle = |{z}\rangle\otimes|{\alpha}\rangle$, and the differentials are
$$
d^2\alpha =\frac{d Re[\alpha] d Im [\alpha]}{\pi} \mbox{ \ \ \ \ and \ \ \ \ \ } d^2 z=\frac{2j+1}{\pi}\frac{d Re[z] d Im[z]}{\left(1+|z|^2\right)^2}.
$$

Expanding the exponential eq. \ref{f} in a Taylor series,
\[
\langle{\alpha,z}| e^{-i \widehat{H} \tau} |{\alpha,z}\rangle = \langle{\alpha,z}| \left[1 - i \widehat{H} \tau + \cdots \right] |{\alpha,z}\rangle,
\]
\begin{equation}
\label{g}\langle{\alpha,z}| e^{-i \widehat{H} \tau} |{\alpha,z}\rangle = 1 - i \langle{\alpha,z}| \widehat{H} |{\alpha,z}\rangle \tau + \frac{(-i\tau )^2}{2!}\langle{\alpha,z}| \widehat{H}^2 |{\alpha,z}\rangle\cdots ,
\end{equation}
where we have used the  normalization of the coherent states  $\langle{\alpha,z}|{\alpha,z}\rangle = 1$. If we assume $\tau<<1$, we  approximate eq. \ref{g} by the first two terms and obtain the leading order semi-classical approximation
\begin{equation}
\label{h}\langle{\alpha,z}| e^{-i \widehat{H} \tau} |{\alpha,z}\rangle \approx 1 - i \langle{\alpha,z}| \widehat{H} |{\alpha,z}\rangle \tau  \approx  e^{-i  \langle{\alpha,z}| \widehat{H} |{\alpha,z}\rangle \tau} = e^{- i H_{cl}\left(\alpha,z \right) \tau}.
\end{equation}

Replacing eq. \ref{h} in eq. \ref{f}, we obtain
\begin{eqnarray}
\nu_Q(E)\approx \nu \left(E\right)& \equiv& \frac{1}{2\pi} \int_{-\infty}^{\infty} \left[\int e^{i E \tau} e^{- i H_{cl}\left(\alpha,z \right) \tau} d^{2}\alpha~d^{2}z \right] d\tau\nonumber\\
 &=& \int \left[\frac{1}{2\pi} \int_{-\infty}^{\infty} e^{i\left(E - H_{cl}\left(\alpha,z \right)\right)\tau} d\tau \right] d^{2}\alpha~d^{2}z\nonumber\\
& =& \int \delta\left[E - H_{cl}\left(\alpha,z \right)\right] d^{2}\alpha~d^{2}z. \label{i}
\end{eqnarray}

If we write equation eq. \ref{i} in canonical variables, we have an interesting interpretation. 
The complex numbers $\alpha$ and $z$ in terms of canonical variables (quadratures for $\alpha$ and a projection over the Bloch sphere for $z$), Eq.(\ref{paraCoh}), are
\begin{equation}
\label{k}\alpha = x+i y=\frac{1}{\sqrt{2}} \left(\frac{q_+}{b} + i\frac{q_-}{c}\right) ~~~ \hbox{y} ~~~ z=X+iY = e^{-i\phi} tan\left(\frac{\theta}{2}\right),
\end{equation}
where  $bc=\hbar$, and  $\theta$ and $\phi$ are the zenith and azimuthal angles of the Bloch sphere. To express the differentials in terms of these variables we need the determinant of the Jacobian matrix,
\begin{eqnarray}
\label{l}d^{2}\alpha &= \frac{d x~d y}{\pi} = \frac{dq_+~dq_-}{\pi} \left|\frac{\partial\left(x,y\right)}{\partial\left(q_+,q_-\right)}\right| ~~~ \hbox{and} \\ \nonumber
d^{2}z &=\frac{2j + 1}{\pi} \frac{d X~d Y}{\left(1 + |z|^{2}\right)^{2}} = \frac{2j + 1}{\pi} \frac{dj_{z}~d\phi}{\left(1 + |z|^{2}\right)^{2}} \left|\frac{\partial\left(X,Y\right)}{\partial\left(j_{z},\phi\right)}\right|.
\end{eqnarray}

For the bosonic variables it is easy 
\begin{equation}
\label{qqq}
\frac{dx~dy}{\pi} = \frac{dq_+~dq_-}{\pi} \left|\frac{\partial\left(x,y\right)}{\left(q_+,q_-\right)}\right| = \frac{dq_+~dq_-}{\pi} \left|
\begin{array}{cc}
\frac{\partial x}{\partial q_+} & \frac{\partial x}{\partial q_-}  \\
\frac{\partial y}{\partial q_+} & \frac{\partial y}{\partial q_-} \end{array}
\right|= \frac{dq_+~dq_+}{2\pi\hbar}
\end{equation}

The calculation for the spin variables is not straightforward. However, if we know how they are projected over the Bloch sphere whose axes are given by $j_{x} = j \cos\phi \sin\theta$, $j_{y} = j \sin\phi \sin\theta$, and $j_{z} = j \cos\theta$, then it is easy to obtain the real ($X$) and imaginary ($Y$) parts of $z$ in terms of the canonical variables $j_{z}$ y $\phi$
\begin{equation}
\label{n}X = \cos\phi\, \sqrt{\frac{j + j_{z}}{j- j_{z}}} ~~~ \hbox{y} ~~~ Y = \sin\phi\, \sqrt{\frac{j + j_{z}}{j- j_{z}}}.
\end{equation}

With this we have
\begin{eqnarray}
\label{o}d^{2}z &= \frac{2j + 1}{\pi} \frac{dX~dY}{\left(1 + |z|^{2}\right)^{2}} = \frac{2j + 1}{\pi} \frac{dj_{z}~d\phi}{\left(1 + |z|^{2}\right)^{2}} |\frac{\partial\left(X,Y\right)}{\partial\left(j_{z},\phi\right)}| = \\ \nonumber
&=\frac{2j + 1}{\pi} \frac{dj_{z}~d\phi}{\left(1 + |z|^{2}\right)^{2}} \left|
\begin{array}{cc}
\frac{\partial X}{\partial j_{z}} & \frac{\partial X}{\partial \phi}  \\
\frac{\partial Y}{\partial j_{z}} & \frac{\partial Y}{\partial \phi} \end{array}
\right|.
\end{eqnarray}

Substituting eq. \ref{n} in eq. \ref{o} in order to calculate the determinant of the Jacobian matrix, we have
\begin{equation}
\label{p}d^{2}z = \frac{2j + 1}{\pi} \frac{dj_{z}~d\phi}{\left(1 + |z|^{2}\right)^{2}} \frac{j}{\hbar \left(j - j_{z}\right)^{2}}. 
\end{equation}

We note that $|z|^{2} = \tan^{2} \left(\frac{\theta}{2}\right) = \frac{j + j_{z}}{j - j_{z}}$. Therefore, eq. \ref{p} is simplified to
\begin{equation}
\label{q}d^{2}z = \frac{2j^{2} + j}{4\pi \hbar j^{2}} dj_{z}~d\phi = \frac{1 + \frac{1}{2j}}{2\pi\hbar}dj_{z}~d\phi. 
\end{equation}

In the limit $j\rightarrow\infty$ we obtain
\begin{equation}
\label{r} d^{2}z = \frac{dj_{z}~d\phi}{2\pi\hbar}.
\end{equation}

Finally, replacing the results eq. \ref{qqq} and eq. \ref{r} in eq. \ref{i} we obtain
\begin{equation}
\label{s}\nu \left(E\right) = \frac{1}{\left(2\pi\hbar\right)^{2}} \int \delta\left[E - H_{cl}\left(\alpha,z\right)\right] dq_+~dq_-~dj_{z}~d\phi.
\end{equation}

This equation represents  the lowest order  semiclassical approximation to the density of sates.


\section{ Minimal energy for the semiclassical Hamiltonian}
We calculate in this appendix the extremal values of the semiclassical Hamiltonian, Eq.(\ref{Hclass}),
\begin{eqnarray}
H_{cl,j}&=\omega_{0}j_{z}+\frac{\omega}{2}\left(q_{+}^{2}+q_{-}^{2}\right)+\frac{\gamma\,j}{\sqrt{\mathcal{N}/2}}\sqrt{1-\left(\frac{j_{z}}{j}\right)^{2}}\times\\ \nonumber
&\times \left[(1+\delta)q_{+}\,\cos(\phi)-(1-\delta)q_{-}\,\sin(\phi)\right].
\end{eqnarray}
Its derivatives are
\begin{eqnarray}
\fl \frac{\partial H_{cl}}{\partial q_{+}}&= \omega\,q_{+} + \frac{\gamma\,j}{\sqrt{\mathcal{N}/2}}\sqrt{1-\left(\frac{j_{z}}{j}\right)^{2}}(1+\delta)\cos(\phi),\\
\fl \frac{\partial H_{cl}}{\partial q_{-}}&= \omega\,q_{-} - \frac{\gamma\,j}{\sqrt{\mathcal{N}/2}}\sqrt{1-\left(\frac{j_{z}}{j}\right)^{2}}(1-\delta)\sin(\phi),\\
\fl \frac{\partial H_{cl}}{\partial j_{z}}&=\omega_{0}-\frac{\gamma}{j\sqrt{\mathcal{N}/2}}\frac{j_{z}}{\sqrt{1-\left(\frac{j_{z}}{j}\right)^{2}}}\left[(1+\delta)q_{+}\,\cos(\phi)-(1-\delta)q_{-}\,\sin(\phi)\right],\\
\fl \frac{\partial H_{cl}}{\partial \phi}&=-\frac{\gamma\,j}{\sqrt{\mathcal{N}/2}}\sqrt{1-\left(\frac{j_{z}}{j}\right)^{2}}\left[(1+\delta)q_{+}\,\sin(\phi)+(1-\delta)q_{-}\,\cos(\phi)\right].
\end{eqnarray} 

From $\partial_{q_{-}}H_{cl}=0$ and $\partial_{q_{+}}H_{cl}=0$, we obtain
\begin{eqnarray}
q_{+}^{m}&=-\frac{\gamma}{\omega}\frac{j}{\sqrt{\mathcal{N}/2}}\sqrt{1-\left(\frac{j^{m}_{z}}{j}\right)^{2}}\left(1+\delta\right)\,\cos(\phi^{m}), \label{Aq+}\\
q_{-}^{m}&=\frac{\gamma}{\omega}\frac{j}{\sqrt{\mathcal{N}/2}}\sqrt{1-\left(\frac{j^{m}_{z}}{j}\right)^{2}}\left(1-\delta\right)\,\sin(\phi^{m})\label{Aq-}.
\end{eqnarray}
Now, from $\partial_{\phi} H_{cl}=0$ we have to possibilities, 
\begin{equation}
j_{z}^{m}=\pm j\,\,\,\,\mbox{or}\,\,\,\,j_{z}^{m}\neq \pm j.
\end{equation}
In the first case, it immediately follows from Eqs.(\ref{Aq+}) and (\ref{Aq-}) that   $q_{-}=0$ and $q_{+}=0$. As $j_{z}^{m}=\pm j$ (north and south poles of the Bloch sphere respectively) the azimuthal angle $\phi$ is completely irrelevant. Then, we have two extremal points for any value of the coupling constant
\begin{equation}
\label{CoorNormal}
(q_{+}^{m},q_{-}^{m},j_{z}^{m},\phi)=(0,0,\pm j,\mbox{undetermined}).
\end{equation}
By evaluating the Hamiltonian at these points we obtain
$$
H(q_{+}^{m}=0,q_{-}^{m}=0,j_{z}^{m}=\pm j,\phi)=\pm\omega_o j.
$$    
The other case, $j_z^m\not=0$, implies, from $\partial_{\phi} H_{cl}=0$ that
\begin{equation}
\label{Aeqn:1}
\left(1+\delta\right)\sin(\phi^{m})q_{+}^{m}=-\left(1-\delta\right)\cos(\phi^{m})q_{-}^{m}.
\end{equation}
On the other hand, from (\ref{Aq+}) and (\ref{Aq-}) it is easy to find
\begin{equation} \label{Aeqn:2}
\left(1-\delta\right)\sin(\phi^{m})q_{+}^{m}=-\left(1+\delta\right)\cos(\phi^{m})q_{-}^{m}.
\end{equation}
In order the two Eqs.(\ref{Aeqn:1}) and (\ref{Aeqn:2}) hold for arbitrary $\delta$ we have the following possibilities:
\begin{enumerate}
\item  $q_{+}^{m}=0$ and $q_{-}^{m}=0$, which gives us the already found extremal points (\ref{CoorNormal}).
\item  $\phi^{m}=0,\pi$, then $\sin(\phi^{m})=0$, $\cos(\phi^{m})=\pm1$, and $q_{-}^{m}=0$ with $q_{+}^{m}\neq 0$. 
\item  $\phi^{m}=\pi/2,3\pi/2$, then $\cos(\phi^{m})=0$, $\sin(\phi^{m})=\pm1$ and $q_{+}^{m}= 0$ with $q_{-}^{m}\neq 0$.
\end{enumerate}
Note that for the Tavis-Cummings case ($\delta=0$) both Eqs. (\ref{Aeqn:1}) and (\ref{Aeqn:2}) become the same,   and the condition holds for every $\phi\in[0,2\pi]$. This case is solved in \cite{Bas14}.

For the second of the latter possibilities, by substituting (\ref{Aq+}) in $\partial_{j_z} H_{cl}=0$, we obtain $(j_z/j)=-(\gamma_{j,+}/\gamma)^2$, where 
\begin{equation}
\gamma_{j,\pm}=\sqrt{\frac{\mathcal{N}}{2j}}\frac{\sqrt{\omega_{0}\omega}}{\left(1\pm\delta\right)}=\sqrt{\frac{\mathcal{N}}{2j}}\,\gamma_{\pm}.
\end{equation}
Then, by substituting this result in (\ref{Aq+}), we obtain $q_+^m$.  The resulting   extremal points are
\begin{equation}
\fl (q_{+}^{m},q_{-}^{m},j_{z}^{m},\phi^{m})=\left(\mp\frac{\gamma (1+\delta)}{\omega}\frac{j}{\sqrt{\mathcal{N}/2}}\sqrt{1-\left(\frac{\gamma_{j,+}}{\gamma}\right)^{4}},0,-j\left(\frac{\gamma_{j,+}}{\gamma}\right)^{2},0\,\,\mbox{or}\,\,\pi\right).
\end{equation}
Since $-j\leq j_z\leq j$, the previous point is valid if and only if $\gamma\geq \gamma_{j,+}$.
By evaluating the Hamiltonian at these points we obtain the energy
$$
E_{j,e}^+=-\frac{j\omega_0}{2}\left[\left(\frac{\gamma_{j,+}}{\gamma}\right)^2+\left(\frac{\gamma}{\gamma_{j,+}}\right)^2\right].
$$

For the third of the previously enumerated possibilities, by substituting (\ref{Aq-})  in $\partial_{j_z} H_{cl}=0$, we obtain $(j_z/j)=-(\gamma_{j,-}/\gamma)^2$. Then, we substitute this result back in (\ref{Aq-}) to obtain  $q_-^m$.  Giving the following  extremal points   
\begin{eqnarray}
&(q_{+}^{m},q_{-}^{m},j_{z}^{m},\phi^{m})=\\ \nonumber		
&\left(0,\pm\frac{\gamma (1-\delta)}{\omega}\frac{j}{\sqrt{\mathcal{N}/2}}\sqrt{1-\left(\frac{\gamma_{j,-}}{\gamma}\right)^{4}},-j\left(\frac{\gamma_{j,-}}{\gamma}\right)^{2},\pi/2\,\,\mbox{or}\,\,3\pi/2\right).
\end{eqnarray}
Which is a valid point, provided that $\gamma\geq\gamma_{j,-}$. The energy of this extremal points is obtained by evaluating the Hamiltonian, the result is
$$
E_{j,e}^-=-\frac{j\omega_0}{2}\left[\left(\frac{\gamma_{j,-}}{\gamma}\right)^2+\left(\frac{\gamma}{\gamma_{j,-}}\right)^2\right].
$$
Clearly, as $\delta\in[0,1]$ it is easy to see that $\gamma_{j,+}\leq\gamma_{j,-}$. We summarize our findings as follows (see equally Table 1 of the main text)
\begin{enumerate}
\item For $\gamma< \gamma_{j,+}$,  the energies of the extremal points  in ascending order are  $-j\omega_o$ (ground-state) and $+j\omega_o$ (local maximum).

\item For $\gamma_{j,+}\leq \gamma <\gamma_{j,-}$, the energies of the extremal points in ascending order are $E_{j,e}^+$ (ground state), $-j\omega_o$ (saddle point) and $+j\omega_o$ (local maximum).

\item For $\gamma_{j,-}\leq \gamma$, the energies of the extremal points in ascending order are  $E_{j,e}^+$ (ground state), $E_{j,e}^-$ (saddle point), $-j\omega_o$ (local maximum) and $+j\omega_o$ (local maximum).
\end{enumerate}


\section{Bounds for the atomic semiclassical variables}
We derive in this appendix analytic expressions for the SDoS, which are determined by the bounds of variables $j_z$ and $\phi$. We follow closely the Appendix A of Ref.\cite{Bas14}. In this reference,   the study was limited to the maximal pseudospin case $j=\mathcal{N}/2$ with   critical  coupling  given by $\gamma_{c}=\gamma_+=\sqrt{\omega\omega_o}/(1+\delta)$. In order to extend the results of that reference to every pseudospin, the following simple substitution has to be made $\gamma_c\rightarrow \gamma_{j,+}=\gamma_+\sqrt{\frac{\mathcal{N}}{2 j}}$.  In the above mentioned appendix,  it was demonstrated that, after integration of the bosonic variables, 
$$
\nu_\delta(E,\mathcal{N},j) =\frac{1}{2\pi \omega}\int dj_z d\phi
$$
and that this integral gives  real values if and only if the following condition is fulfilled
\begin{equation}
\left(\frac{\gamma}{\gamma_{j,-}}\right)^{2}\sin^{2}(\phi)+\left(\frac{\gamma}{\gamma_{j,+}}\right)^{2}\cos^{2}(\phi)\geq\frac{2\left(y-\epsilon_{j}\right)}{\left(1-y^{2}\right)},
\end{equation}
where  $y=j_z/j$, $\epsilon_j=E/(\omega_o j)$, and we have  used $\gamma_{j,\pm}$ defined in Eq.(\ref{defgamm}). The previous condition determines the bounds of variables $j_z$ and $\phi$. To see more clearly these bounds we rewrite the previous  condition as
\begin{eqnarray}
\label{condi}
&\cos^{2}(\phi)\geq g_j(y,\epsilon_j) {\mbox{\ \ \ with \ \ \ }}\\ \nonumber
&g_j(y,\epsilon_j)\equiv\left(\left(\frac{\gamma}{\gamma_{j,+}}\right)^{2}-\left(\frac{\gamma}{\gamma_{j,-}}\right)^{2}\right)^{-1}\left[\frac{2\left(y-\epsilon_{j}\right)}{\left(1-y^{2}\right)}-\left(\frac{\gamma}{\gamma_{j,-}}\right)^{2}\right].
\end{eqnarray}
Clearly, if $g_j(y,\epsilon_j)<0$ the condition is satisfied for every $\phi\in[0,2\pi)$, whereas if   $g_j(y,\epsilon_j)>1$ the condition is never fulfilled. For $0\leq g_j(y,\epsilon_j)\leq 1$ the condition is fulfilled for two intervals of $\phi$, $[-\phi_l,\phi_l]$ and $[\pi-\phi_l,\pi+\phi_l]$, with 
\begin{equation}
\label{limphi} 
\phi_l=\arccos\left[\sqrt{g_j(y,\epsilon_j)}\right].
\end{equation}   Therefore, to determine the bounds of the variables $\phi$ and $y=j_z/j$, it is necessary to study the behavior of function $g_j(y,\epsilon_j)$ in the interval $y\in[-1,1]$ for different couplings and energies. 

\begin{figure}
\centering{
\begin{tabular}{c c}
\multicolumn{2}{c}{a)}\\
\multicolumn{2}{c}{\includegraphics[angle=0,width=0.5\textwidth]{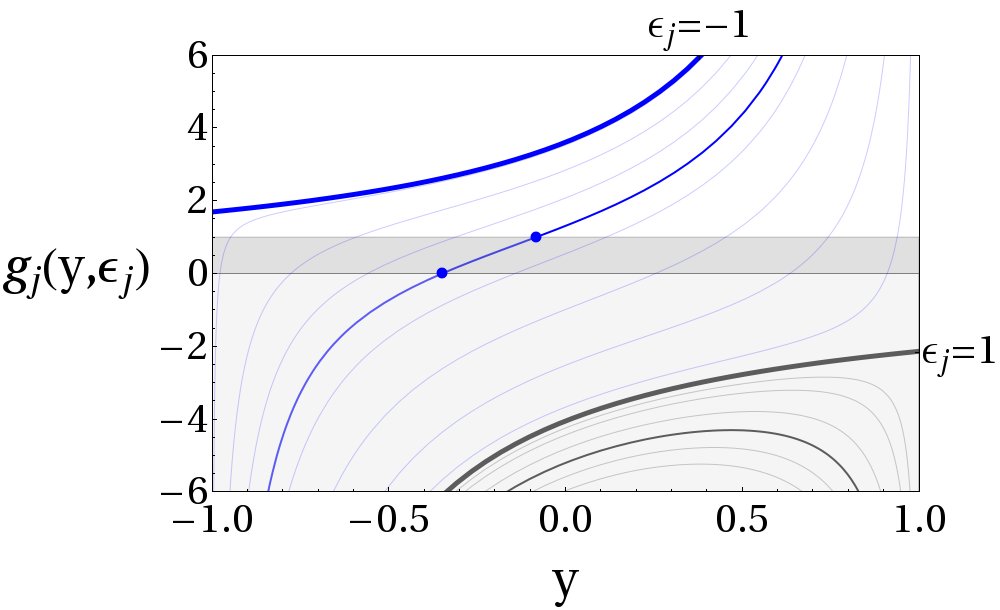} }\\
b)&c)\\
\includegraphics[angle=0,width=0.5\textwidth]{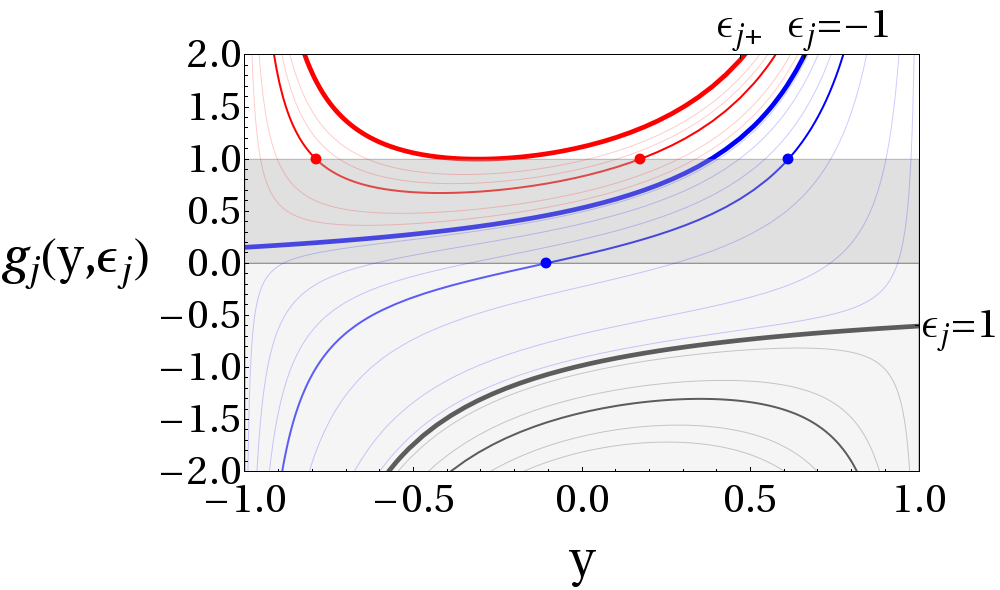} &\includegraphics[angle=0,width=0.45\textwidth]{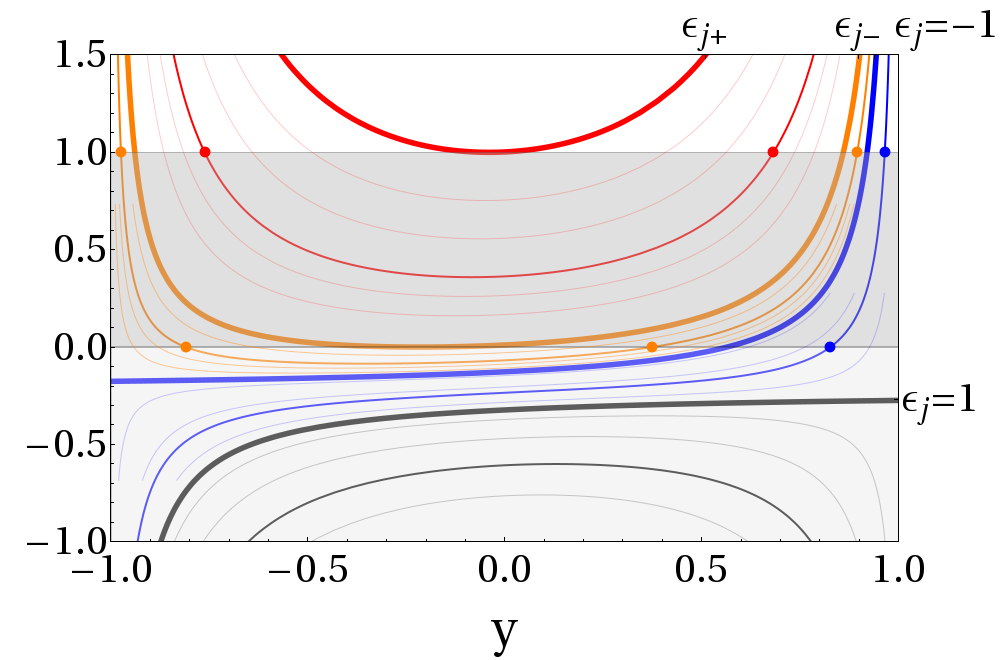} \\
\end{tabular}}
\caption{ Function $g_j(y\epsilon_j)$ as a function of $y=j_z/j$, for three different couplings $\gamma/\gamma_{j,+}=0.8$, $1.8$ and $5.0$ [panels a) b) y c) respectively],  for different energies $\epsilon_j=E/(\omega_o j)$. The couplings  correspond, respectively, to the three different coupling regimes a) $\gamma<\gamma_{j,+}$, b) $\gamma_{j,+}\leq\gamma<\gamma_{j,-}$ and c) $\gamma_{j,-}\leq \gamma $. In each panel, curves with similar behavior are plotted  with the same color.  The   thickest lines, indicate  critical curves  separating different curve behaviors, the corresponding critical energies are indicated at  right and top frame boundaries.  In  selected curves, the crossings with the horizontal lines at  values $0$ and $1$ are indicated by solid circles, these crossings define, respectively,  the roots $y_{0\pm}$ and $y_{1\pm}$ discussed in the text. The gray zones indicate the region where the condition (\ref{condi}) is fulfilled,  in the light gray zones the condition is satisfied by whatever value of angle $\phi$ in the whole  interval $[0,2\pi)$, whereas in the dark gray zones the condition is satisfied only by bounded values of $\phi$.    A   $\delta=0.4$ was used.}
\label{fig:10}
\end{figure}

The different  behaviors of function  $g_j(y,\epsilon_j)$ are  classified according to three couplings intervals $[0,\gamma_{j,+})$, $[\gamma_{j,+},\gamma_{j,-})$ and $[\gamma_{j,-},\infty)$. For each of these coupling intervals, different energy regimes can be identified.  In figure \ref{fig:10} a graphical summary of the behavior of function   $g_j(y,\epsilon_j)$ is presented. Three different coupling were selected representing the three coupling intervals mentioned above. 

\subsection{$\gamma<\gamma_{j,+}$}
For the first coupling, $\gamma=0.8\gamma_{j,+}\in[0,\gamma_{j,+})$, the function  $g_j(y,\epsilon_j)$ takes values below to $1$ only for $\epsilon_j\geq -1$. In the energy interval $-1\leq\epsilon_j<1$ [blue lines in panel a) of Fig.\ref{fig:10}], the function is less than $0$ (implying that  $\phi$ can take values from $0$ to $2\pi$) only for $-1\leq y \leq y_{0+} $, where $y_{0+}$ is the largest   root of 
$$
g_j(y,\epsilon_j)=0,
$$
whose roots  are given by 
$$
y_{0\pm}=-\left(\frac{\gamma_{j,-}}{\gamma}\right)^{2}\pm\left(\frac{\gamma_{j,-}}{\gamma}\right)\sqrt{2\left(\epsilon_{j}-\epsilon_{j-}\right)},
$$
with $\epsilon_{j-}$ defined by 
$$
\epsilon_{j\pm}=-\frac{1}{2}\left[\left(\frac{\gamma_{j,\pm}}{\gamma}\right)^{2}+\left(\frac{\gamma}{\gamma_{j,\pm}}\right)^{2}\right].
$$
For $y_{0+}< y\leq y_{1+}$, with $y_{1+}$ defined below, the function $g_j(y,\epsilon_j)$ takes values in the interval $[0,1]$, consequently the angle $\phi$  is limited to two intervals around $0$ and $\pi$, as  explained before  Eq.(\ref{limphi}).  The value $y_{1+}$ is the largest  root of equation
$$
g_j(y,\epsilon_j)=1,
$$
with roots  given by
$$
y_{1\pm}=-\left(\frac{\gamma_{j,+}}{\gamma}\right)^{2}\pm\left(\frac{\gamma_{j,+}}{\gamma}\right)\sqrt{2\left(\epsilon_{j}-\epsilon_{j+}\right)}.
$$
For this energy interval, the values $y>y_{1+}$ are forbidden, and  the density of states is given by
\begin{eqnarray}
\fl 2\pi\omega\nu_\delta (E,\mathcal{N},j)&=\int d j_z d\phi = j \int dy  d\phi =\\ \nonumber
&= j \int_{-1}^{y_0+}  2 \pi dy+ j\int_{y_{0+}}^{y_{1+}}  (\phi_l+\phi_l)+ \left (\pi+\phi_l-(\pi-\phi_l)\right) dy \nonumber\\
&= 2\pi j (y_{0+}+1)+ 4j \int_{y_{0+}}^{y_{1+}} \arccos\sqrt{g_j(y,\epsilon_j)} dy.     \nonumber      
\end{eqnarray}

For $\epsilon_j>1$ [black lines in panel  a) of Fig.\ref{fig:10}], the function $g_j(y,\epsilon_j)<0$ for all $y\in [-1,1]$, consequently neither  $\phi$ nor $j_z$ are limited (the whole Bloch sphere become available). Hence, the density of states is
$$
2\pi\omega\nu_\delta (E,\mathcal{N},j)= \int d j_z d\phi=2j 2\pi=4\pi j.
$$ 


\subsection{$\gamma_{j,+}\leq \gamma <\gamma_{j,-}$}

In this case, corresponding to panel b) of Fig.\ref{fig:10} ($\gamma=1.8 \gamma_{j,+}$), the two energy intervals of the previous case \{$\epsilon_j\in [-1,1)$ and $\epsilon_j\in [1,\infty)$ \} remain together the corresponding expressions for the density of states. But a new energy interval appears $\epsilon_{j+}\leq \epsilon_j <-1$. For this energy interval, the  function $g_j(y,\epsilon_j)$ [red lines in panel b) of Fig. \ref{fig:10}] takes values  less than $1$  and greater than $0$,  in the interval  $y_{1-}\leq y \leq y_{1+}$, with  the roots $y_{1\pm}$ defined above. Consequently, the angle $\phi$ is bounded  as  explained before  Eq.(\ref{limphi}). Hence, the density of states for this energy interval is
$$
2\pi\omega\nu_\delta (E,\mathcal{N},j)= \int d j_z d\phi = 4 j \int_{y_{1-}}^{y_{1+}} \arccos\sqrt{g_j(y,\epsilon_j)} dy.
$$    


\subsubsection{$\gamma_{j,-}\leq \gamma$.}

Finally, the  behavior of $g_j(y,\epsilon_j)$ for this case is depicted in panel c) of Fig. \ref{fig:10} ($\gamma=5\gamma_{j,+}$). The energy intervals and corresponding expression for the density of states are similar to the previous cases with the following changes: the interval $\epsilon_j\in [\epsilon_{j,+},-1)$ changes to $[\epsilon_{j,+},\epsilon_{j,-})$, and a new intertwined interval emerges, that given by $\epsilon_{j,-}\leq \epsilon_j<-1$. In this latter interval,  the function  $g_j(y,\epsilon_j)$ [orange lines in panel c) of Fig. \ref{fig:10}] takes values in the interval $[0,1]$ for $y\in[y_{1-},y_{0-}]$ and $y\in [y_{0+},y_{1+}]$, where  the values of $\phi$ are bounded by Eq.(\ref{condi}). On the other hand,  for $y\in[y_{0-},y_{0+}]$ the function $g_j(y,\epsilon_j)$ is less than $0$ and, consequently,  there the angle $\phi$ can take values in the whole interval $[0,2\pi)$. Therefore, for this new energy  interval, $\epsilon_j\in [\epsilon_{j,-},-1)$, the density of states is
\begin{eqnarray}
\fl &2\pi\omega\nu_\delta (E,\mathcal{N},j)=\\ \nonumber
\fl &=4 j \int_{y_{1-}}^{y_{0-}} \arccos\sqrt{g_j(y,\epsilon_j)} dy+4 j \int_{y_{0+}}^{y_{1+}} \arccos\sqrt{g_j(y,\epsilon_j)} dy +2\pi j (y_{0+}-y_{0-}).
\end{eqnarray}

\end{document}